\documentclass[12pt,preprint]{emulateapj}

\def \apj {ApJ}

\def \apjl {ApJ}
\def \solphys {Solar Phys.}

\def \aap {A\&A}

\begin{document}

\title{Spectropolarimetric observations of the \ion{Ca}{2} 8498 \AA\ and 8542 \AA\ lines in the quiet Sun}

\author{A. Pietarila\altaffilmark{1}, H. Socas-Navarro}
\affil{High Altitude Observatory, National Center for Atmospheric Research\altaffilmark{2}, 3080 Center Green, Boulder, CO 80301, USA\vbox{}}
\author{T. Bogdan}
\affil{Space Environment Center, National Oceanic and Atmospheric Administration, 325 Broadway, Boulder, CO 80305, USA\vbox{}}

\altaffiltext{1}{Institute of Theoretical Astrophysics, University of Oslo, P.O.Box 1029 Blindern, N-0315 Oslo, Norway}
\altaffiltext{2}{The National Center for Atmospheric Research (NCAR) is sponsored by the National Science Foundation.}

\begin{abstract}

The \ion{Ca}{2} infrared triplet is one of the few magnetically
sensitive chromospheric lines available for ground-based
observations. We present spectropolarimetric observations of the 8498
\AA\ and 8542 \AA\ lines in a quiet Sun region near a decaying active region and compare the results
with a simulation of the lines in a high plasma-$\beta$ regime. Cluster
analysis of Stokes $V$ profile pairs shows that the two lines, despite
arguably being formed fairly close, often do not have similar shapes. In the
network, the local magnetic topology is more important
in determining the shapes of the Stokes $V$ profiles than the phase of the wave, contrary to what our simulations show. We also find that
Stokes $V$ asymmetries are very common in the network, and the histograms of
the observed amplitude and area asymmetries differ significantly from the simulation. Both the network and internetwork show oscillatory behavior in the \ion{Ca}{2} lines. It is stronger in the network, where shocking
waves, similar to those in the high-$\beta$ simulation, are seen and
large self-reversals in the intensity profiles are common.

\end{abstract}

\section{Introduction}

Our understanding of solar magnetic fields outside active regions
has increased significantly during the last years. This is due to new
and better instrumentation ({\em e.g.}, {\em THEMIS}, \citealt*{Paletou+Molodij2001}; {\em VSM} on {\em SOLIS}, \citealt*{Keller2001}; {\em Swedish Solar Telescope}, \citealt*{Scharmer2003}; {\em Solar Optical Telescope} on {\em Hinode}, \citealt*{Shimizu2004}; and {\em SPINOR}, \citealt*{Socas-Navarro+others2006a}), better diagnostic techniques (see for example \citealt*{BellotRubio2006} for a review on inversion techniques) and advanced numerical simulations
(\citealt*{Stein+Nordlund2006} and references therein). A large portion of the work has focused on photospheric magnetic
fields. Only now
we are starting to have adequate tools for investigating chromospheric magnetism in more detail. (For a review of chromospheric magnetic fields see \citet{Lagg2005}). This is
not surprising considering the numerous difficulties in observing
chromospheric magnetic fields, interpreting the data, and performing
realistic MHD simulations.  

There are two different sets of lines that are often used
for chromospheric spectropolarimetry, the \ion{He}{1} infrared (IR)
triplet at 10830 \AA, and the \ion{Ca}{2} IR triplet at 8500 \AA. Both
line sets have their advantages and disadvantages. The \ion{He}{1}
lines are formed over a relatively thin layer, and therefore
observations can be inverted using a simple Milne-Eddington model. The
drawback is that while the formation range is fairly narrow, the
precise formation height remains uncertain, and the Milne-Eddington
inversions do not give any information on the atmospheric
gradients. The lines are also sensitive to the Paschen-Back effect,
which must be included in the inversion code
\citep{Socas-Navarro+TrujilloBueno+LandiDeglInnocenti2004}. Furthermore,
simulating the \ion{He}{1} lines is difficult since coronal irradiation has a
non-negligible effect on their formation \citep{Andretta+Jones1997}. In contrast, the formation of
the \ion{Ca}{2} IR lines is fairly well understood \citep{Lites+Chipman+White1982}. The broad
\ion{Ca}{2} lines sample a large region of the atmosphere, from the
photosphere to the lower chromosphere. However, the \ion{Ca}{2} lines
are formed in nonLTE, making inversions considerably more cumbersome.

Several investigations using the \ion{Ca}{2} IR lines have studied intensity
and velocity oscillations in the quiet Sun
(e.g. \citealt{Lites+Chipman+White1982, Deubner+Fleck1990}) or, alternatively, magnetic
fields in active regions
(e.g. \citealt{Socas-Navarro+others2000b}). In both cases the
lines have proven useful as diagnostics of the solar chromosphere. In this paper we present results
of spectropolarimetric observations of two of the lines in an enhanced
network region. We have both spatial maps and time series data. The
observations show that the \ion{Ca}{2} lines are formed in a very
interesting region, namely the region where the atmosphere is
transforming from a plasma dominated ($\beta >> 1$) to a magnetic field dominated ($\beta << 1$)
regime in terms of dynamic force balance. Wave propagation is clearly seen in the
highly dynamic magnetic regions, whereas the weakly magnetic internetwork is found to
be less variable. Interestingly, the two \ion{Ca}{2} lines
exhibit significant differences even though in calculations they are formed fairly close
together. The importance of gradients in the chromospheric network is
clearly demonstrated by the prevalence of asymmetric Stokes $V$
profiles in the data.

The paper is arranged as follows: in \S\ \ref{sec:data} the data and
their reduction are addressed. Results of analyzing the data using different approaches are presented in \S\
\ref{sec:results}. We performed cluster analyses on the Stokes $V$
profiles to classify them and to describe spatial patterns seen in the
data. Statistics, such as profile amplitudes and asymmetries, are
presented. The time dependent behavior of the lines in different
network and internetwork regions is also discussed. In \S\
\ref{sec:highb} the observations are compared to simulations of the
lines in a high plasma-$\beta$ regime (\citealt{Pietarila+others2006},
hereafter P06). Finally, in \S\ \ref{sec:end} the main results are
summarized and discussed.

\section{Observations and data reduction}
\label{sec:data}
\begin{figure*}
\epsscale{1}
\plotone{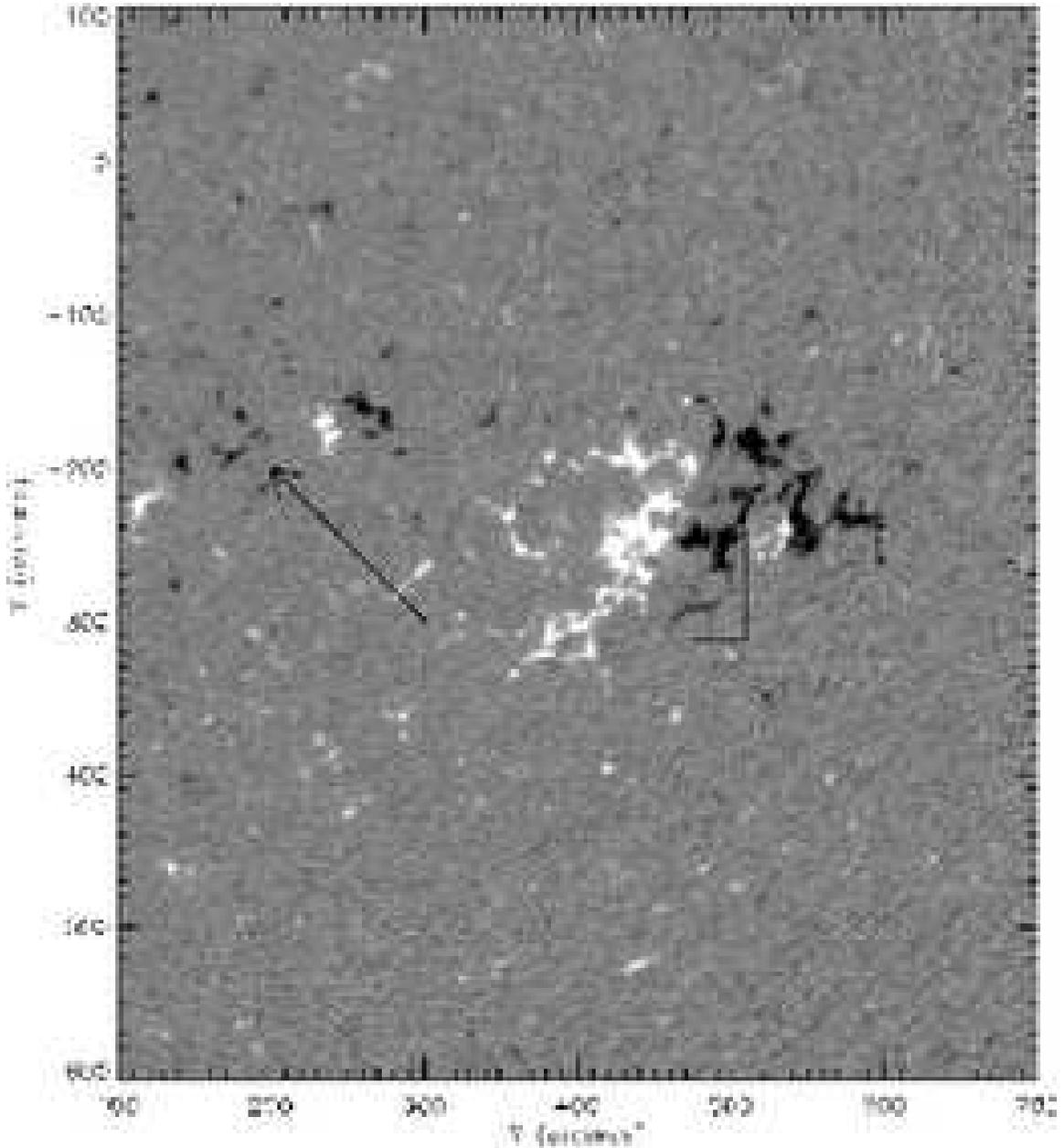}
\caption{MDI magnetogram showing the position of the slit for the time series and the map (rectangular region). The observed region was close to the decaying active region, AR10763. 
\label{fig:MDI}
}
\end{figure*}

The Spectro-Polarimeter for INfrared and Optical Regions (SPINOR,
\citealt{Socas-Navarro+others2006a}) at the Dunn Solar telescope,
Sacramento Peak Observatory, was used to observe two of the
\ion{Ca}{2} infrared triplet lines at 8498 \AA\ and 8542 \AA, as well
as two photospheric \ion{Fe}{1} lines at 8497 \AA\ and 8538 \AA. The setup
included several other lines but because of computer problems
only data from the two Ca lines which used the ASP TI TC245 cameras
were recorded fully. The data have 256 points in both the wavelength
and spatial position with a typical noise level of $6\times 10^{-4}$ $I_c$ (1 $\sigma$ deviation from the mean)
and a spectral sampling of 25 m\AA. The pixel height corresponds to
$\approx$ 0.38 arcseconds on the solar surface along the slit. We observed a quiet
Sun region near disk center at S17.3 W32.1 on May 19, 2005 at 14:14
UT. An MDI-magnetogram of the region is shown in Figure
\ref{fig:MDI}. The slit was positioned in the vicinity of a decaying
active region, AR10763, but avoided flux concentrations from the active region (i.e., plages). A time series consisting of 99 time steps of
short scans (3 slit positions), with a spacing of 0.375 arcseconds
each, was acquired during variable seeing conditions. The cadence is
$\approx$ 10 seconds (i.e., a given slit position was repeated every 30
s). The time series was followed by a 63 step raster centered around
the position where the slit was during the time series. The raster
step size was 0.375 arcseconds. Adaptive optics (AO,
\citealt{Rimmele2000}) were used during the observing sequence but the
compromised seeing conditions did not allow for continuous locking onto
granulation. This caused the slit to jump occasionally, making the longest period
with a stationary slit in the time series 17 time steps (8.5 min). The
spatial resolution varied during the sequences being at best less than
an arc second, but on average a factor of two worse.

Standard procedures for flat field and bias were used for the data
reduction. Instrumental polarization was removed using the available calibration
data, as explained in \citet{Socas-Navarro+others2006a}. No absolute
wavelength calibration was attempted because no suitable telluric lines are
present. Instead a wavelength calibration using spatial pixels devoid
of magnetic field was done by fitting the average spectrum to the Kitt
Peak FTS-spectral atlas \citep{Neckel+Labs1984a}. The FTS atlas was
also used to find the normalization factor for the intensities to the
quiet Sun continuum intensities. Because of detector flatfield residuals and prefilter shape, the continua in the raw data from both detectors are
tilted. The tilts were removed {\it a posteriori} by subtracting a linear fit ($y=a+b\lambda$) obtained by matching the continuum intensity
levels to those of the FTS atlas.

 The data were analyzed using both the raster and time series for
statistical purposes. The period when the slit was
stationary on the solar disk was used to study the time-dependent behavior of the lines. Because of the short length of this period, we do not present any Fourier analysis of the data. 
To make a classification of Stokes V profile morphologies, we did 
cluster analyses based on a Principal Component Analysis (PCA) in a similar manner as the work of  \citet{SanchezAlmeida+Lites2000} and \citet{Khomenko+others2003} for
photospheric lines. We
computed amplitudes for Stokes $I$ and $V$ profiles. Because the Ca
line intensity profiles often exhibit strong self-reversals, no
proxies for atmospheric velocities, such as lines' centers of
gravity or bisectors, are adequate. For those Stokes V profiles with amplitudes greater than $7 \times 10^{-3}$ $I_c$,
(i.e. $\ge 10\sigma$), amplitude and area
asymmetries were also calculated.

The amplitude asymmetry of a Stokes V profile is defined by \citep
{MartinezPillet+others1997}:

\begin{eqnarray}
\sigma_{\mathrm{a}}= \frac {a_b-a_r}{a_b+a_r},
\end{eqnarray}
where $a_b$ and $a_r$ are the unsigned extrema of the blue and red
lobes of the Stokes $V$ profile.

The area asymmetry of a Stokes $V$ profile is defined by \citep {MartinezPillet+others1997}:

\begin{eqnarray} 
\label{eqn:area}
\sigma_{\mathrm{A}}=s\frac{\int_{\lambda_0}^{\lambda_1} V(\lambda)d\lambda}{\int_{\lambda_0}^{\lambda_1} |V(\lambda)|d\lambda},
\end{eqnarray}
where $s$ is the sign of the blue lobe. Because of the broad, deep lines and large velocities (compared with the photosphere) present in the chromosphere, the
choice of the integration range for the area asymmetries is
non-trivial for the Ca lines. We followed the same procedure as in
P06. In the weak field regime the Stokes V profile is
proportional to $dI/d\lambda$ (strictly true only in the absence of
atmospheric velocity and magnetic gradients). Inspection of the data showed that most
of the observed Stokes $V$ profiles have roughly the same structures as the $dI/d\lambda$ profiles. The intensity
in the blue wing ($\lambda_0$) of the line profile was matched with a point in the
red wing ($\lambda_1$) with the same intensity. The signal-to-noise in the intensity
profiles is much higher than in the Stokes V profiles and also
the slope is much steeper. This makes matching points with the same
value more accurate in the intensity than in the Stokes V
profiles. The selection of a wavelength to start the integration range
 was made by choosing a wavelength point that is far enough from
the line core so that self-reversals are not an issue. In our data this point,
$\lambda_1$, is at 600 m\AA\ from reference wavelength of line center. The same value was used in P06.

\begin{figure*}
\epsscale{0.5}
\plotone{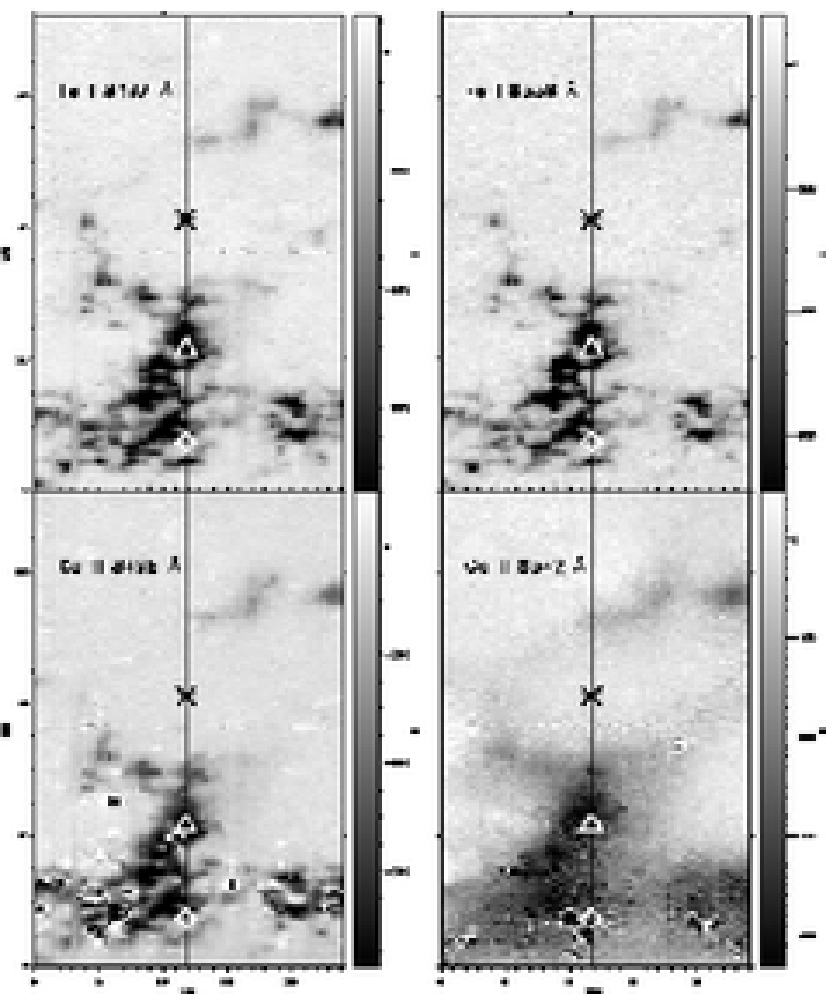}
\caption{Magnetograms of the map deduced by using the weak field method \citep{LandiDeglInnocenti1992}. The Stokes $V$ signal is measured in units of  Gauss. Vertical lines show the position of slit during the time series. First panel:  \ion{Fe}{1} 8497 \AA, second panel: \ion{Fe}{1}8538 \AA, third panel: \ion{Ca}{2} 8498 \AA, third panel: \ion{Ca}{2} 8542 \AA. Location on the solar disk: S32.1, W17.3.  The orientation
of the magnetograms is 180 degrees from the MDI magnetogram in Figure \ref{fig:MDI}. The plotted symbols ($\ast$, $\Diamond$ and $\bigtriangleup$) on the images show where the pixels discussed later in the text are located.
\label{fig:magnetogram}
}
\end{figure*}

Magnetograms made from the 63 step scan are shown in Figure
~\ref{fig:magnetogram}. The panels are in order of increasing
formation height: \ion{Fe}{1} 8497 \AA, \ion{Fe}{1} 8538 \AA,
\ion{Ca}{2} 8498 \AA\ and \ion{Ca}{2} 8542 \AA. The lower part of the
slit was located above a flux concentration along the enhanced network and the upper part
over an internetwork region with very little magnetic
activity. The network becomes wider and more diffuse with increasing line formation height as described by \citet{Giovanelli1980}. Not all magnetic flux seen in the
photosphere can be identified in the chromosphere and vice versa. However, interpreting the chromospheric magnetograms is difficult due
to the self-reversed features in the cores of the Ca line
Stokes $V$ profiles. 
 
\section{Results}
\label{sec:results}
\begin{figure*}
\epsscale{1} \plotone{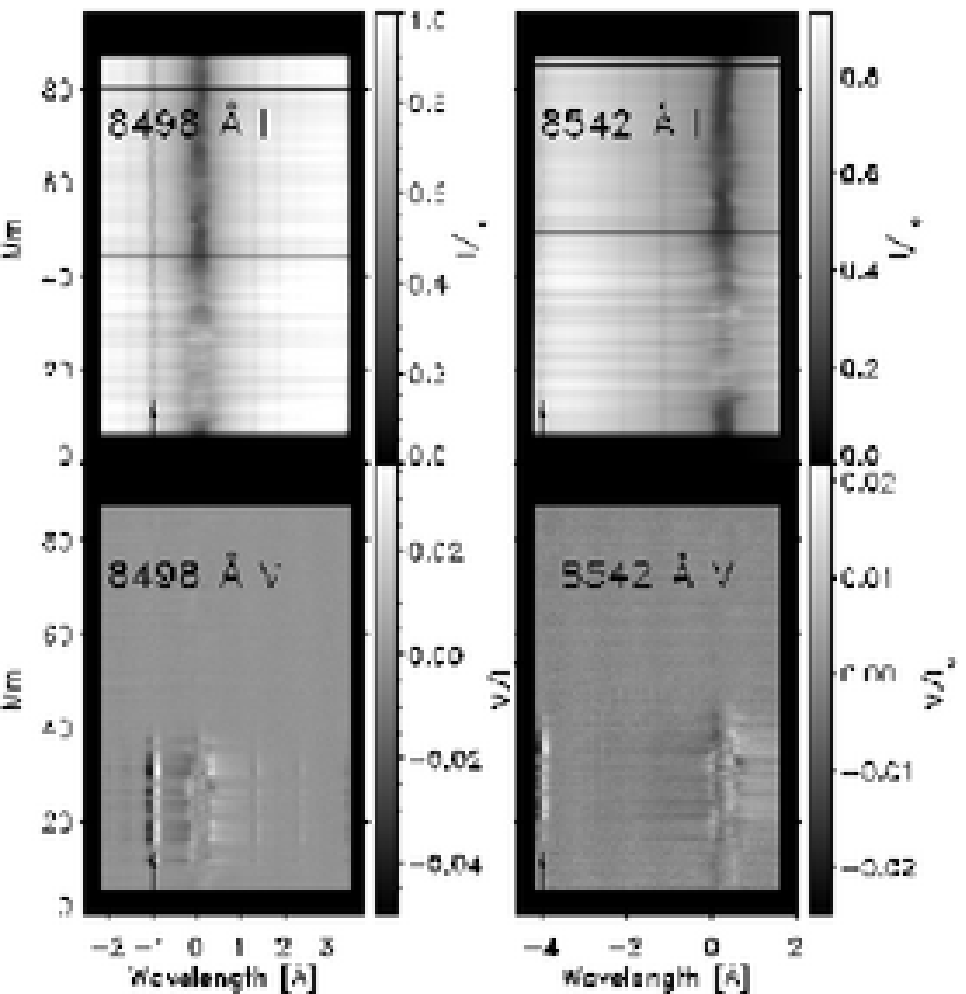}\caption{Dispersed images of the slit. The arrows mark the locations of the two photospheric iron lines and the horizontal lines in the intensity images are the hairlines used to
spatially coalign the two detectors. Wavelengths are measured from 8498 \AA\ (left) and 8542 \AA\ (right). 
\label{fig:slits}
}
\end{figure*}

In Figure ~\ref{fig:slits}, Stokes $I$ and $V$ spectra of the solar surface under the slit are
shown for both \ion{Ca}{2} lines as well as the two photospheric \ion{Fe}{1} lines in the Ca lines' wings (marked by arrows). Since the \ion{Fe}{1} 8497 \AA\ line
is blended in the Ca line's wing and the \ion{Fe}{1} 8538 \AA\ line is
very close to the edge of the detector, no quantitative analysis is
done for them. No signal above the noise was recorded in Stokes $Q$ and $U$
so they will not be addressed in what follows. Residual vertical fringing
caused by the polarization modulator is visible in the Stokes $V$ images. We chose not
to try to remove the fringing since its' amplitude is of the same
order of magnitude as the noise.

The network, present in the lower part of the slit, is associated with less absorption in
the intensity profiles. Both Ca lines often show self reversals, which
are usually stronger on the blue side of the line than in the red. The Stokes $V$
profiles of both Ca lines have large, extended wings. At times, the profiles may have both
polarities present on the blue side of the core but in almost all
cases the far blue wing of the profile has the same polarity (i.e., opposite sign) as the red
wing.  The Stokes $I$ and $V$ profiles of the chromospheric lines look
distinctively different from the photospheric lines: the Ca lines have
more structure, they are wider and exhibit more spatial variation than
the photospheric Fe lines. Some differences are seen between the two
Ca lines: the 8542 \AA\ line is slightly broader, has more structure
in the spectra and also stronger absorption than the 8498 \AA\ line.

The internetwork region, present in the upper part of the slit, is mostly devoid of
Stokes $V$ signal, and Stokes $I$ is more homogeneous than in the
network. Self reversals are usually not seen in the profiles. A small
portion of the internetwork region has structures in Stokes $I$ that
are similar to those seen in the magnetic region: Stokes $I$ is
brighter than in the surrounding areas and the profiles show some self
reversals. Closer inspection of the images reveals a visible, albeit a
very small amplitude, Stokes $V$ signal.

\begin{figure*}
\epsscale{0.5} \plotone{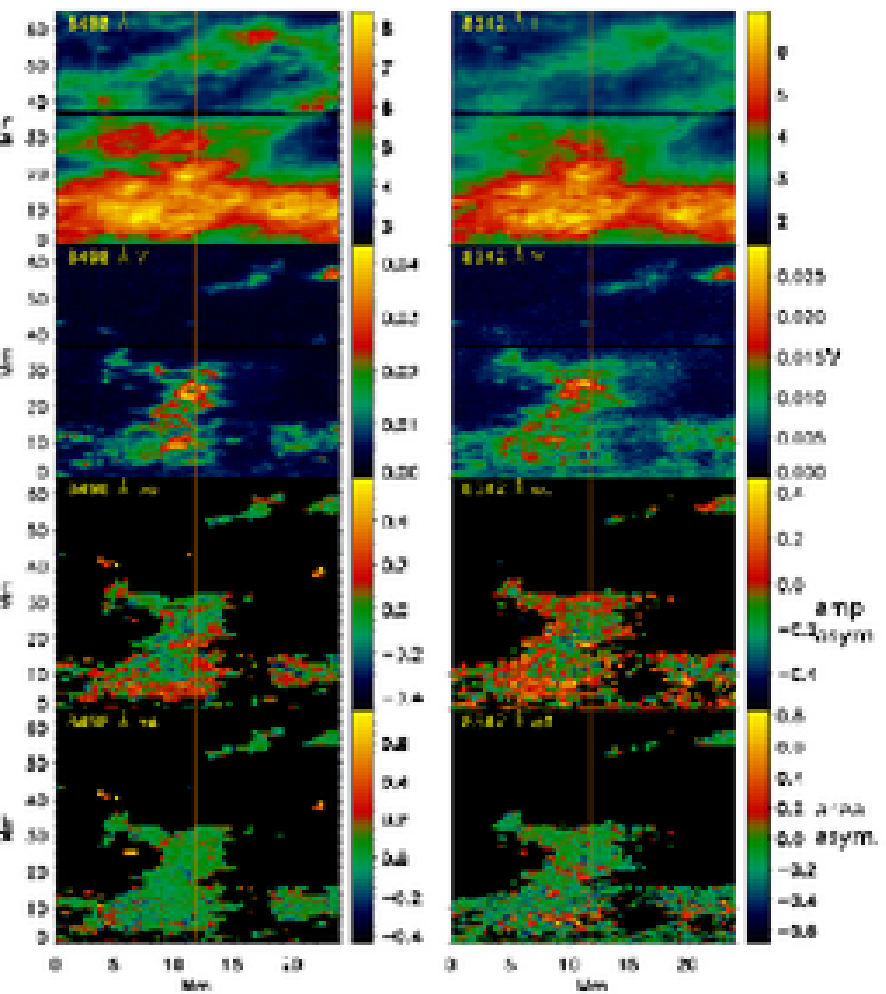}
\caption{Maps of the 8498 \AA\ and 8542 \AA\ lines' Stokes $I$ amplitudes, Stokes $V$ amplitudes, area asymmetries and amplitude asymmetries in the raster scan. The horizontal line seen in the amplitude images is a hairline used to spatially coalign the detectors. The vertical line shows the position of the slit during the time series. Note that x-axis is stretched compared with y-axis.
\label{fig:map-image}
}
\end{figure*}

The spatial patterns of Stokes $I$ and $V$ amplitudes and asymmetries
in the two Ca lines (Figure \ref{fig:map-image}) are fairly similar to one another.
The network is clearly visible in the Stokes $I$ and $V$ amplitudes, though it is more diffuse in
the 8542 \AA\ line. There is a structure in the upper part of the map
that is seen best in the 8498 \AA\ intensity image. Parts of this
structure appear also in both lines' Stokes $V$ amplitude and
asymmetry images. The edges of the network have more asymmetric Stokes $V$ profiles. This is seen clearly in the 8542 \AA\
amplitude asymmetry. 

Photospheric velocities can be estimated from the locations of the iron lines' intensity minima. Except for a nearly constant offset caused by the convective down flows in the network, the internetwork and network regions have very similar spatial and temporal patterns. 

\subsection{Classification of the Stokes V profiles}

To classify the shapes of the 8498 \AA\ and 8542 \AA\ Stokes $V$ profile pairs we used PCA, \citep{Rees+others2000} and cluster analysis. The cluster analyses were performed separately for the map and the time series. Here we present a
summary of the PCA procedure and cluster analysis for completeness.


With the PCA we are able to reduce the number of parameters needed to
describe a given profile. Each profile, $S(\lambda_j)$,
$j=1,...,N_{\lambda}$ ($N_{\lambda}$ is the number of wavelength
points in the profile) is composed of a linear combination of
eigenvectors $e_i(\lambda_j)$, $i=1,...,n$:

\begin{eqnarray}
S(\lambda_j)=\Sigma^n_{i=1} c_ie_i(\lambda_j), \label{eq:PCA}
\end{eqnarray}
where the $c_i$ are appropriate constants. The eigenvectors and constants
for a given set of profiles are obtained from a singular value
decomposition (SVD, \citealt{Rees+others2000}, \citealt{Socas-Navarro+others2001}) and form an orthonormal basis with $N_{\lambda}$ eigenvectors:

\begin{eqnarray}
\Sigma^{N_{\lambda}}_{j=1}e_i(\lambda_j)e_k(\lambda{_j})=\delta_{ik}.
\end{eqnarray}

Not all eigenvectors contain the critical information needed to reproduce the
profiles, some of the eigenvectors carry information about the noise
pattern. We can therefore truncate the series expansion and use only a
small number of eigenvectors and corresponding coefficients to
reproduce a given profile. The PCA guarantees that when expansion of
Eq. \ref{eq:PCA} is truncated at a given order $m$, the amount of information in the lower orders is maximized.

We performed the SVD for the two 8498 \AA\ and 8542 \AA\ Stokes $V$ profiles separately. The resulting
orthonormal bases, and also the cluster analysis, depend on the subset of profiles used to construct
it. Because of this we included all Stokes $V$ profiles from pixels where the 8498 \AA\ Stokes $V$ amplitude is above $7\times10^{-3}I_c$, altogether 13671 profiles. Visual inspection of the
eigenvectors shows that the first 11 eigenvectors (approximately)
contain relevant information about the actual shape of the profiles
whereas the remainder are associated with the noise patterns.

The Stokes $V$ profile pairs, now described
with 11 $\times$ 2 coefficients corresponding to the 11 $\times$ 2
eigenvectors instead of 102 (51 for each profile) wavelength points, were organized into a predefined
number of clusters. Before doing this the vectors consisting of the 22 coefficients were standardized, i.e., no information of the absolute Stokes $V$ amplitudes is left, only the relative amplitudes of the 8498 \AA\ and 8542 \AA\ profiles. Based on the values of the coefficients, 6 cluster
centers were identified using the $k$-means method \citep{MacQueen1967}. It starts with $k$ random clusters, which through iterations are changed to minimize the variability within a cluster and maximize it between clusters. Each profile pair is then assigned to the nearest
cluster center in the 22-dimensional Euclidean space. The choice of number of clusters
used for the cluster analysis is non-trivial. Since each data point is
described by 22 numbers we cannot visually distinguish patterns in the
spatial distribution of the points. Instead the number of clusters was
defined by trial and error, i.e. so that each profile type in the time
series or map is represented and each cluster is still clearly distinct
from one another.  For each cluster a profile was constructed using the
eigenvectors and the averaged 2 $\times$ 11 coefficients of all
profiles belonging to that cluster. 

Cluster analysis of the map shows the shapes of Stokes V profiles in network regions with different magnetic
topologies, whereas the time series analysis describes how a set of profiles
from a certain magnetic topology changes with time.

\begin{figure*}
\epsscale{1} \plotone{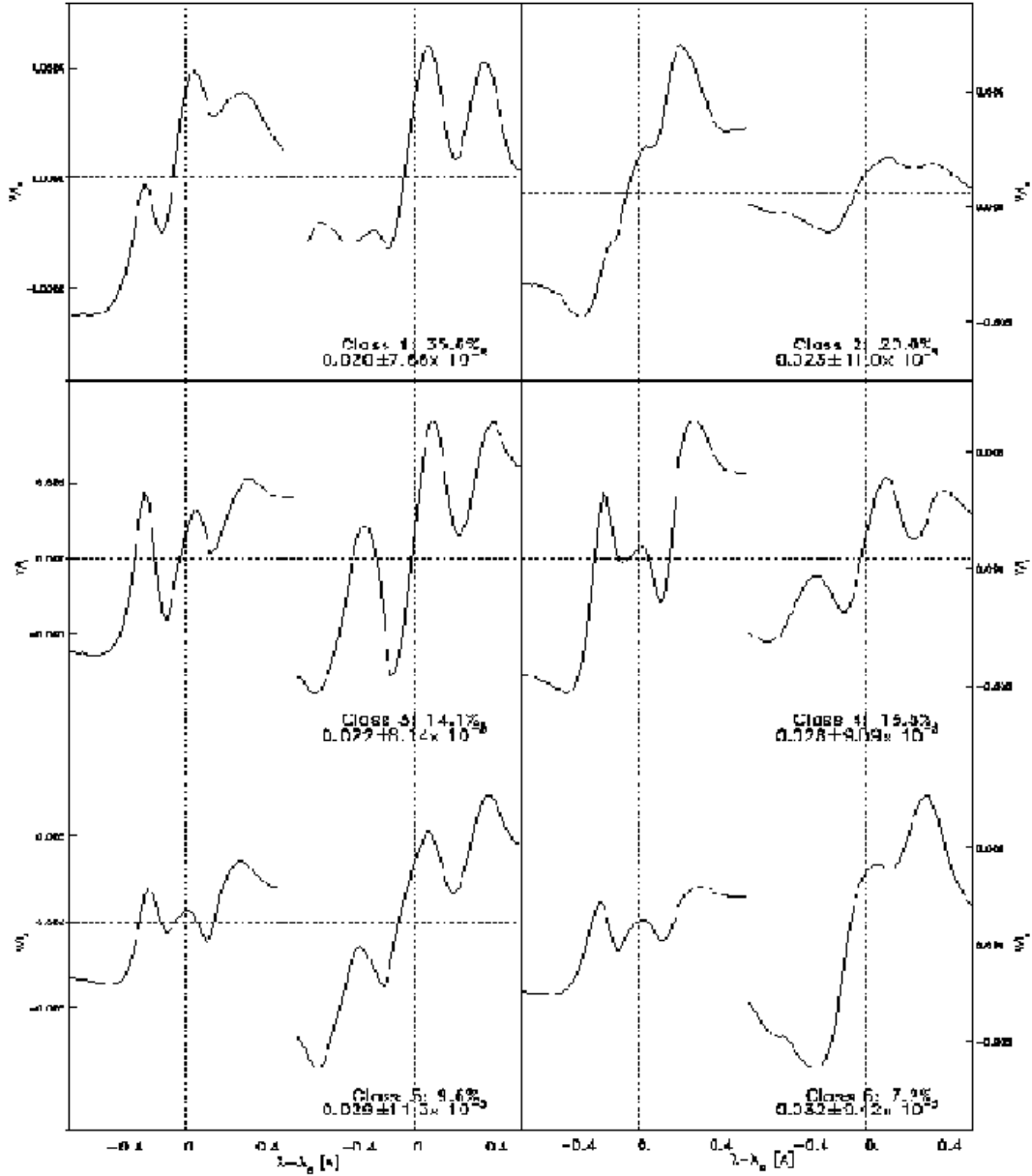}
\caption{Results of cluster analysis of the Stokes $V$ profile pairs in the map. Line on left is 8498 \AA\ and on right 8542 \AA. Shown are the percentages of profile pairs belonging to each cluster, and the mean distance and its standard deviation of the profiles to the cluster center.
\label{fig:clust_map}
}
\end{figure*}

The results for the map are shown in Figure
~\ref{fig:clust_map}. Above each profile is the percentage of all
profiles belonging to the cluster, the mean distance in the Euclidean space of the profiles to
the cluster center, and the standard deviation of the mean. The smaller
the distance to the cluster center, the more compact the cluster is
and the better the cluster describes the profiles. The standard
deviation is proportional to the spread of the distances in each cluster. In general,
clusters with the least number of profiles belonging to them have
larger mean distances.

Three points can be deduced from the figure. First, asymmetric profiles
should be common. In fact, they appear to be more common than symmetric ones. Second,
even though the two Ca lines are formed fairly close to one another
(the 8498 \AA\ line core optical depth is unity at about 1 Mm and the
8542 \AA\ 0.2 Mm higher up in the radiation hydrodynamic simulations
 by Carlsson and Stein 1997), the 8498 \AA\ and 8542 \AA\
profiles in a given cluster are often clearly different from one
another. Third, in all cluster profiles the far-red wings
have the same polarity as the far-blue wings, indicating that the lower parts along the line-of-sight of the
atmosphere, where the wings are formed, are dominated by a single
magnetic polarity.

\nocite{Carlsson+Stein1997}

\begin{figure*}
\epsscale{1} \plotone{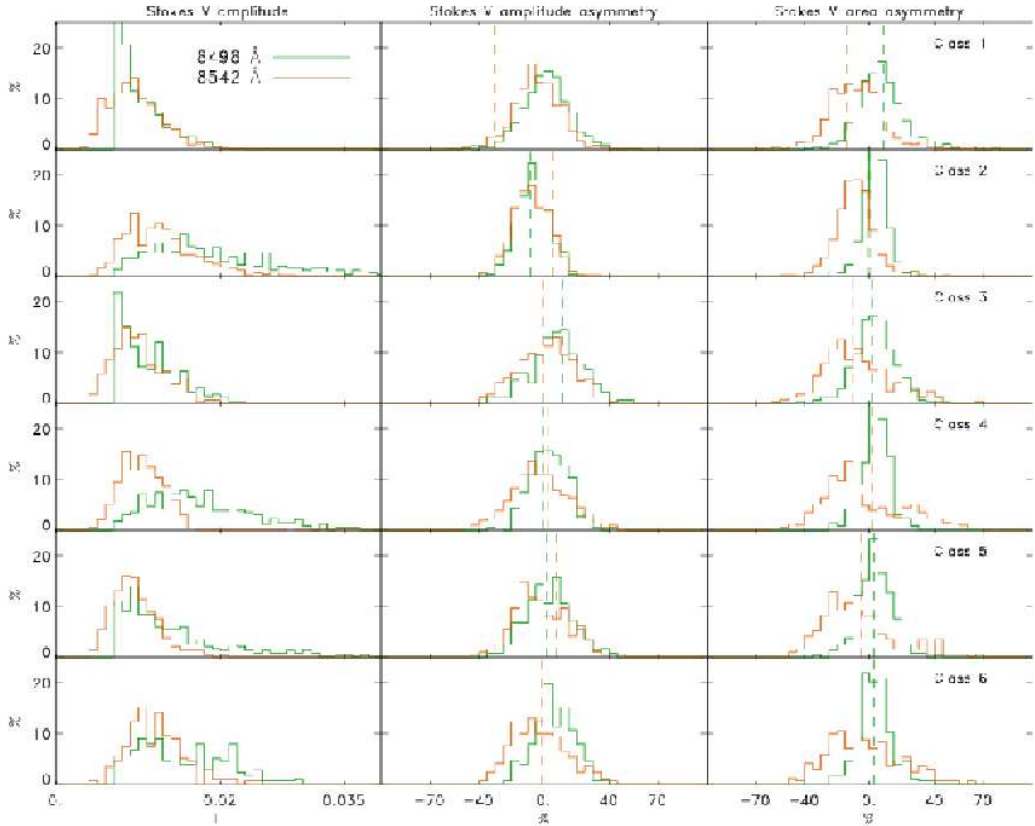}
\caption{Stokes $V$ statistics of the map clusters. The histograms are for all profiles belonging to the given cluster and the dotted vertical lines show the area and amplitude asymmetries for the cluster profiles. 
\label{fig:cstat_map}
}
\end{figure*}

\begin{figure*}
\epsscale{1.} \plotone{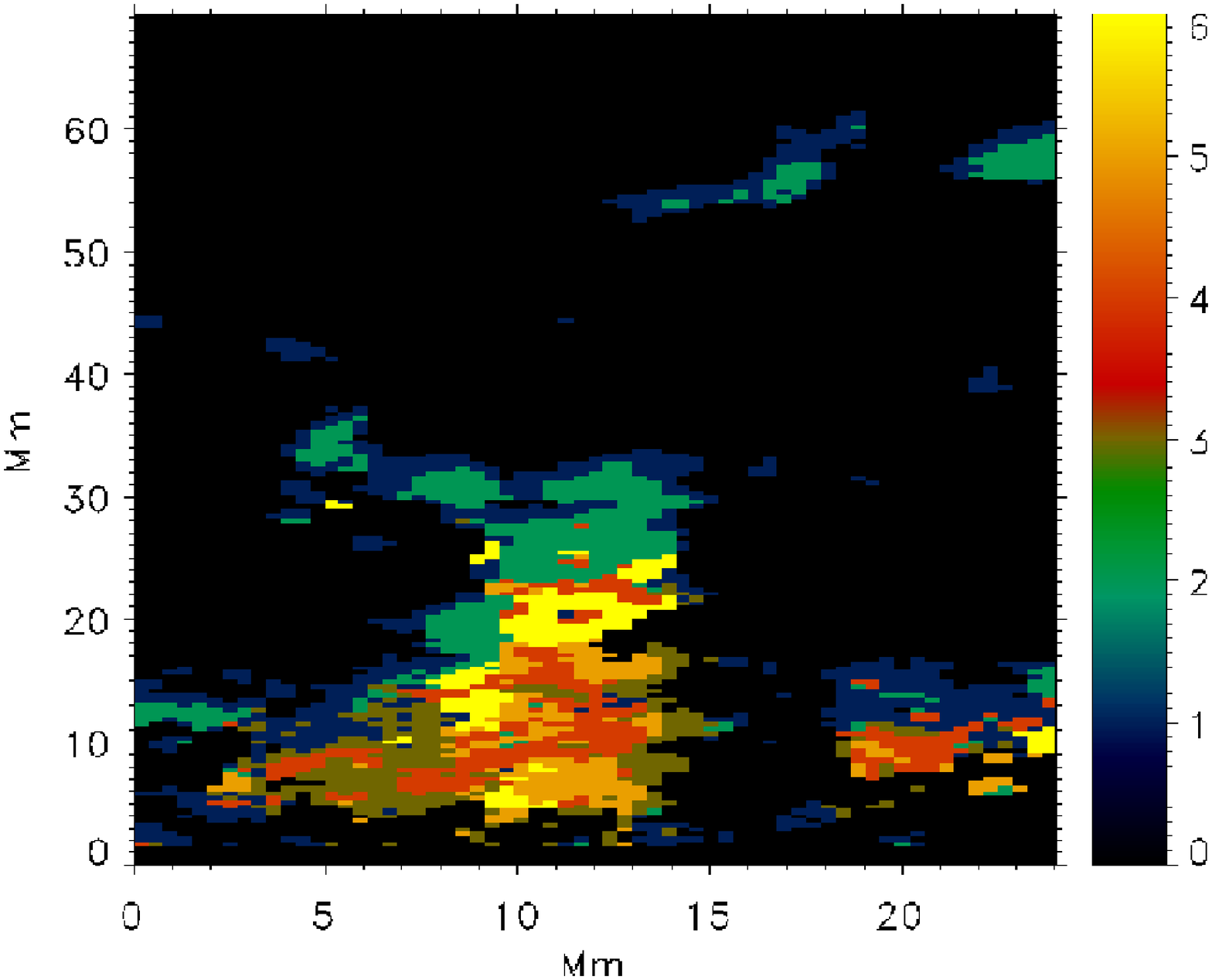}
\caption{Spatial distribution of the clusters in the map. The black areas (0 cluster) correspond to regions where the Stokes $V$ amplitudes are below $7\times10^{-3} I_c$ and where no cluster analysis was performed.
\label{fig:cmap}
}
\end{figure*}

The clusters differ from one another in several different ways: the degree of
asymmetry, and distinct relationships between the 8498 \AA\ and
8542 \AA\ line profiles, relative amplitudes, etc. However,
quantitative measures, such as profile asymmetries, of the clusters do not necessarily represent the members of a
given cluster very well. For example, the variation of Stokes $V$ amplitude
asymmetries within a cluster is large and the mean is not necessarily
the same as that of the cluster profile. The cluster analysis retrieves
qualitative similarities and gives a basis for morphological
classification, rather than representing quantitative similarities
within the data. To illustrate this point, Figure~\ref{fig:cstat_map} displays histograms of the clusters showing the Stokes $V$
amplitudes and asymmetries for all profiles belonging to a given
cluster. Shown in Figure ~\ref{fig:cmap} is the spatial distribution of
the clusters. The smallest network patches often consist of
only cluster 1 and cluster 2 profiles. The middle of the largest network
patch is a mixture of different clusters.

In most cases, the profiles at the edges of the network patches belong
to cluster 1. This is the most common cluster consisting of 35.6
\% of all the profile pairs in the map. The cluster 1 profiles are asymmetric, 8542
\AA\ more so than 8498 \AA, and they also have opposite signs of
amplitude and area asymmetries. The amplitude histograms of profiles
belonging to this cluster show that they have in general low amplitudes,
as one might expect from profiles located at the edges of the
network. The large amplitude asymmetry in the 8542 \AA\ cluster profile
is not seen in the observed profiles. In fact, only very few profiles
exhibit such large asymmetries and there is only a slight tendency of
the profiles having more often negative than positive amplitude
asymmetries. The cluster area asymmetries are in better agreement with
the observed profiles belonging to this cluster.

Regions of cluster 2 profiles are often located adjacent to patches of cluster 1 profiles. The cluster 2 profiles account for 20.0 \% of all profile pairs in the map. The cluster profiles are fairly antisymmetric. This is seen in
the observed profiles as well: the asymmetry histograms tend to be
narrow and only slightly offset from zero. The relative amplitudes of
the two cluster profiles are very different: the 8498 \AA\ amplitude is
a factor 3 larger. The disproportionality is not as large in the
observed profiles though the amplitude histograms show that in general
8498 \AA\ has a larger amplitude than 8542 \AA. The
range of observed amplitudes is considerably larger than in cluster 1.

Of the profile pairs in the map 14.1 \% belong to cluster 3. Also, these profiles are often found in regions close
to the network edges by the patches of cluster 1 profiles. Both cluster
profiles have multiple lobes and are asymmetric, 8498 \AA\ more in
amplitude and 8542 \AA\ in area. This is also seen in the histograms
of the observed asymmetries. There is a strong emission feature on the
blue side of the line in the 8498 \AA\ cluster profile. It is weaker in the 8542
\AA\ profile. The histograms for cluster 3 are nearly identical to those
of cluster 1. This illustrates how cluster analysis based on PCA is captures the qualitative differences in the line profiles.

Cluster 4 consists of 13.5 \% of the profile pairs. Most of the observed
profiles belonging to this cluster are near to the middle of
the largest network patches. The 8498 \AA\ cluster profile is
dominated by a strong emission feature in the blue lobe. This feature
is not visible in the 8542 \AA\ cluster profile. The overlap between the
two lines' amplitude histograms is fairly small. Also the cluster
profiles show this difference in the relative amplitudes: 8542 \AA\ has a
significantly lower amplitude than 8498 \AA. Except for the 8542 \AA\
area asymmetry histogram, all histograms are centered around zero. The
range of area asymmetries in the 8542 \AA\ line is large and the
distribution is skewed towards negative values. This trend in the 8542
\AA\ area asymmetries is seen in several of the clusters.
 
The patches of profiles belonging to the fifth cluster (9.6 \%) are also
found in the less homogeneous middle regions of the network
elements. The 8498 \AA\ cluster profile has a factor 2 lower amplitude
than 8542 \AA. This is not seen in the amplitude histograms but there
is a large overlap between the two histograms. The cluster profiles are
fairly antisymmetric and also the histograms of observed profile asymmetries are
centered around zero. The 8542 \AA\ area asymmetry is again the
exception: it is centered around a negative value.

Cluster 6 is the smallest cluster with 7.2 \% of the profiles. Patches of cluster 6 profiles are located in regions with cluster 4
and 5 profiles. The 8498 \AA\ cluster profile is very similar to that of
cluster 5. Like cluster 5, the 8542 \AA\ cluster 6 profile has a factor 2
larger amplitude and the amplitude histograms overlap nearly
entirely. All the cluster 6 histograms are very similar to cluster 5. The
major difference between the two is that there is very little
structure in the 8542 \AA\ line profile.

\subsection{Time-dependent behavior}

\begin{figure*}
\epsscale{1} \plotone{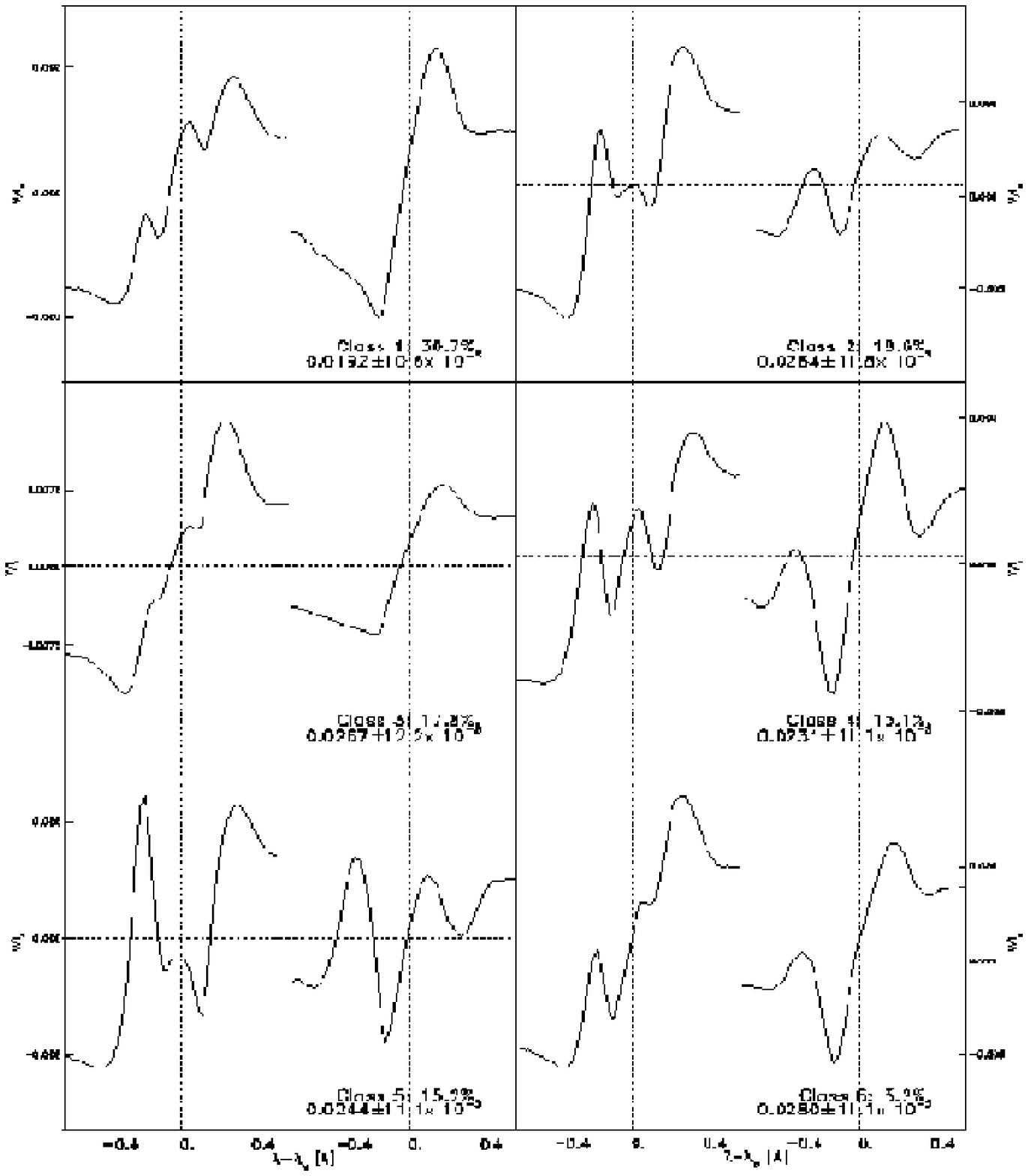}
\caption{As Fig. \ref{fig:clust_map} but for the time series.
\label{fig:clust_ts}
}
\end{figure*}

\begin{figure*}
\epsscale{1} \plotone{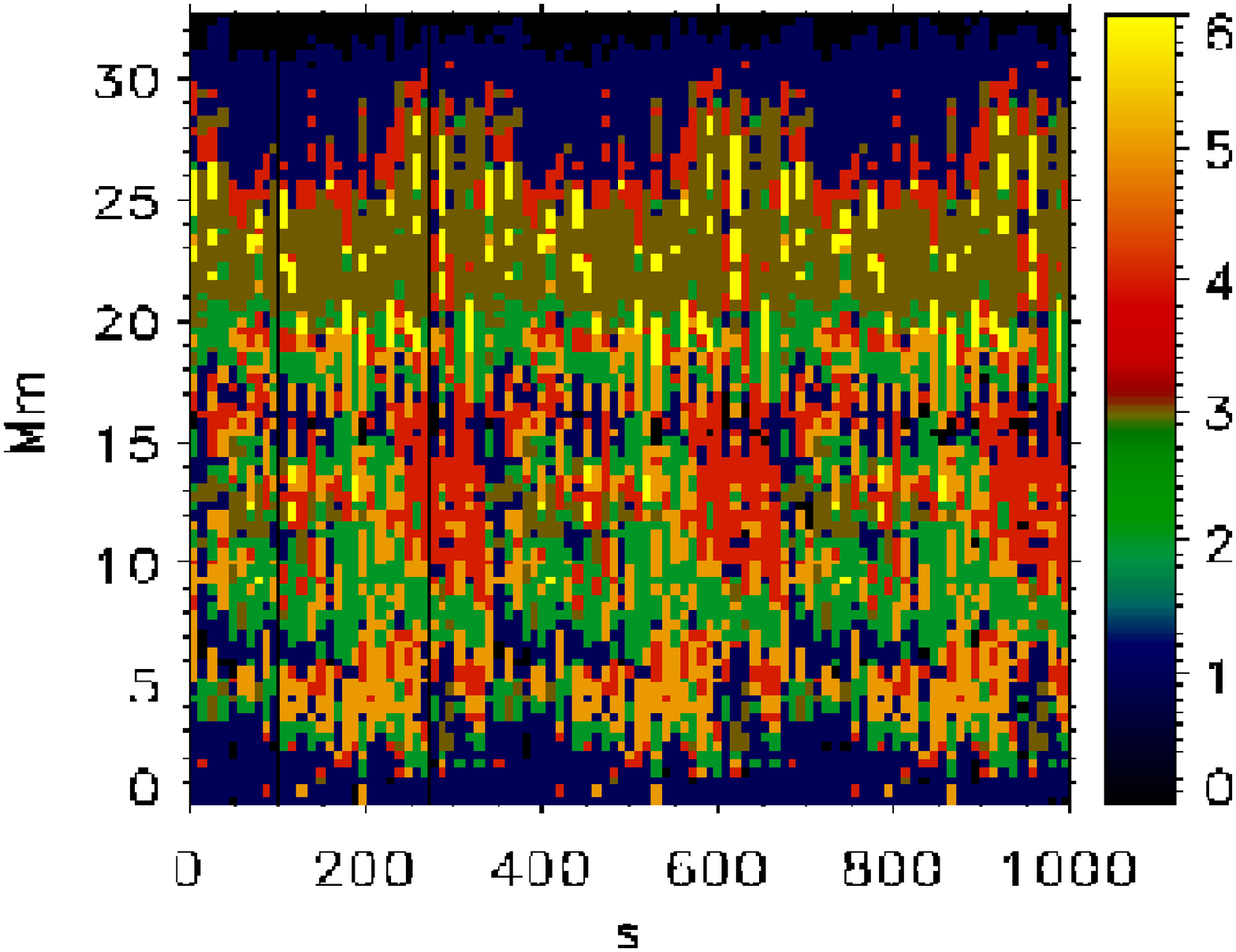}
\caption{The spatio-temporal distributions of the clusters for the first slit position in the time series. The black areas correspond to regions where the Stokes $V$ amplitudes is $\le 7 \times 10^{-3}I_c$. The vertical lines show the period with the best seeing when the slit was stationary.
\label{fig:tsmap}
}
\end{figure*}

The cluster analysis results of the time series are shown in Figure
~\ref{fig:clust_ts}, and the spatio-temporal distribution is captured in Figure
~\ref{fig:tsmap}. The clusters consist of profiles at rest with
varying degrees of structure, and profiles where the blue side is in
emission. While there are temporal changes in the clusters, there are no clear
periodic patterns visible. Most slit positions have a preferred cluster or in some cases the slit position is dominated by two clusters. Positions where more than 2 clusters are dominant are rare.

Because the slit moved occasionally during the time series, no meaningful power spectra can be
made from this data set. The time series data do however allow for a
qualitative analysis of the time-dependent behavior. Comparing network
and internetwork pixels reveals some interesting features: the
network, especially in the intermediate flux regions, is very dynamic
with propagating shock-like features and large self-reversals
appearing frequently in both Stokes $I$ and $V$. In comparison, the internetwork is less dynamic,
intensity oscillations are present but they are much weaker than in the
network. No structures indicating the presence of shocks, are seen in
the internetwork profiles. In agreement with prior observations of chromospheric
lines (e.g., \citealt{Noyes1967}), any oscillation periods in the network
appear to have a longer period  than in the internetwork.

We now examine three different regions, namely an internetwork
pixel, an intermediate flux network pixel, and a strong network pixel.

\subsubsection{Internetwork}

\begin{figure*}
\epsscale{1} \plotone{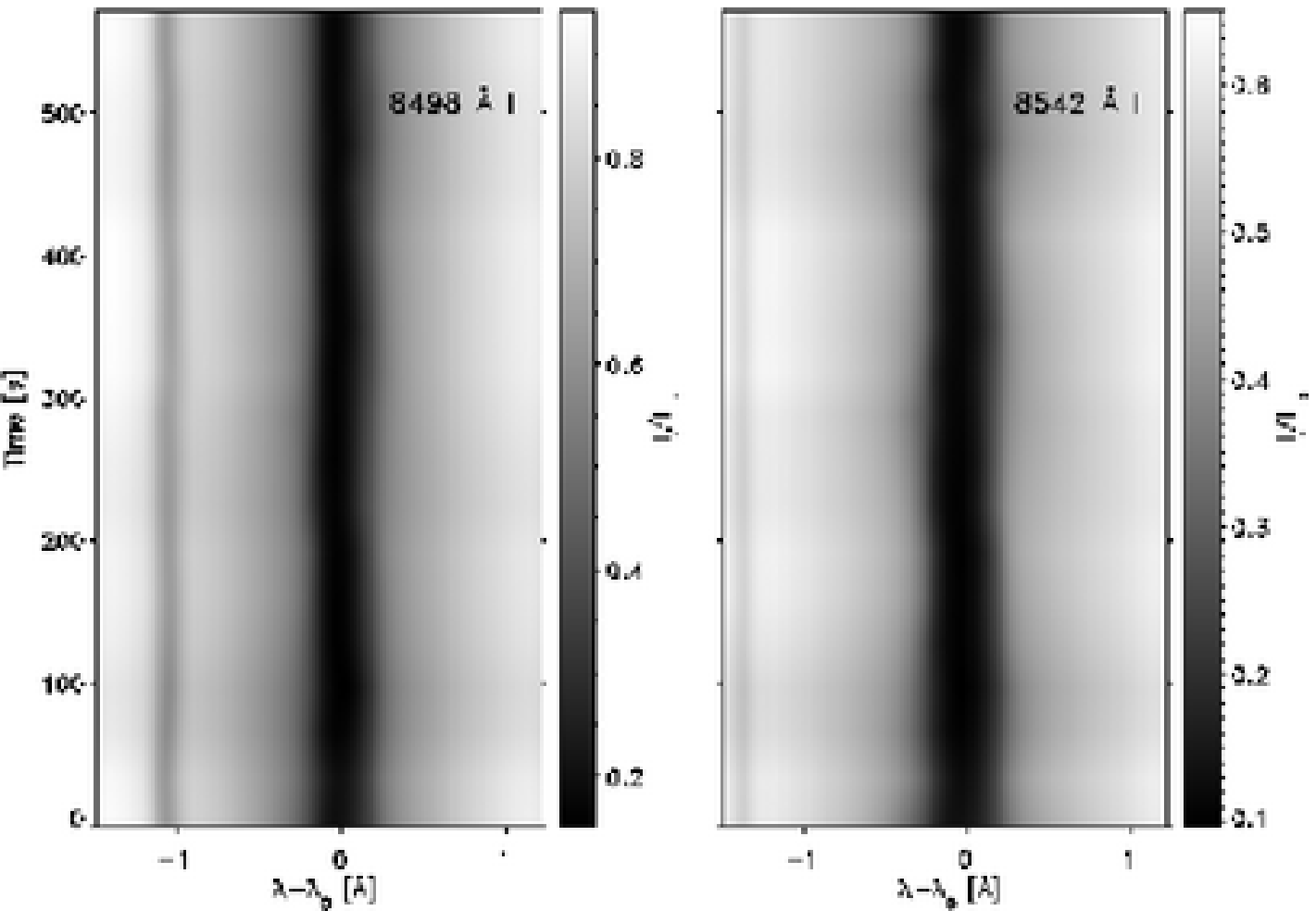}
\caption{Time dependent behavior of Stokes $I$ in an internetwork pixel.  Location of the pixel is marked with an asterisk in
Figure \ref{fig:magnetogram}.
\label{fig:image130}
}
\end{figure*}

In Figure ~\ref{fig:image130} the time evolution of a typical internetwork
pixel is shown. The location of the pixel is marked by an asterisk in
Figure \ref{fig:magnetogram}. The data were taken when the slit was
stationary. No Stokes $V$ signal above the
noise level is seen in the pixel. The Stokes $I$ profiles of both Ca lines change periodically in width and position of the line center, but no self-reversals are
seen. Also, the line-wing intensity shows some oscillations. 

\subsubsection{Intermediate flux network}

\begin{figure*}
\epsscale{1} \plotone{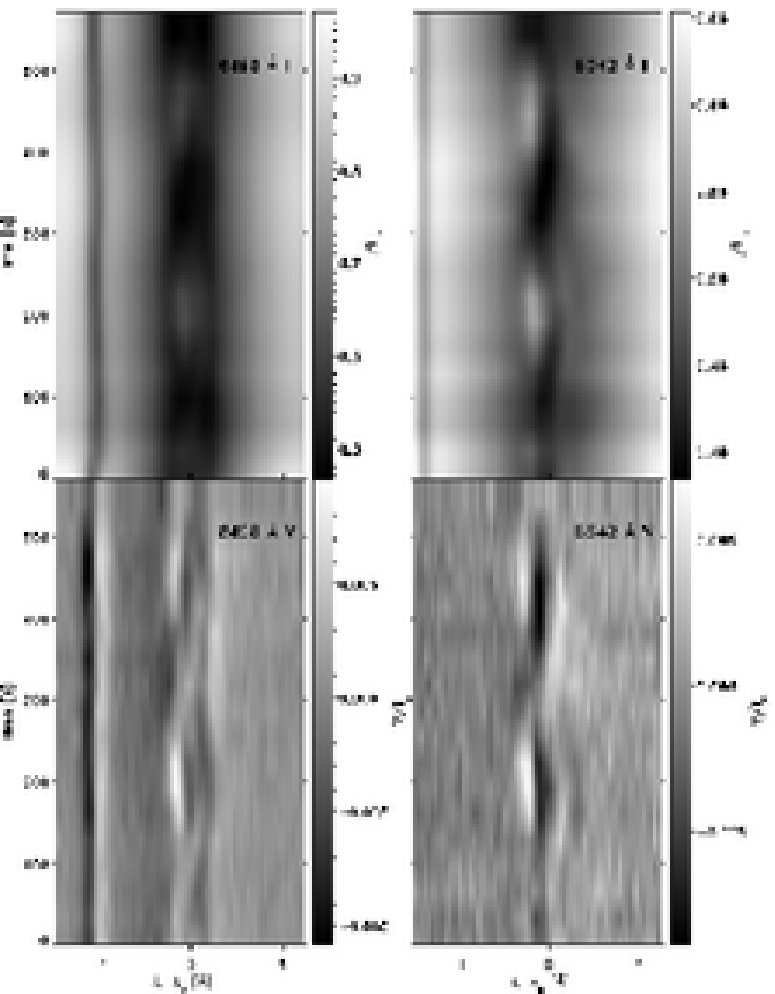}
\caption{Time dependent behavior of Stokes $I$ and $V$  in an intermediate flux pixel. Location of the pixel is marked with a diamond in
Figure \ref{fig:magnetogram}.
\label{fig:image40}
}
\end{figure*}

The difference between the internetwork and network regions with
intermediate flux (Fig. \ref{fig:image40}) is dramatic: the network
region is much more dynamic, and highly asymmetric profiles, in both
lines Stokes $I$ and $V$, are seen. The time dependent behavior of the
photospheric iron line is quite similar to what is seen in the
internetwork. 

The Stokes $I$ in both lines has a clearly oscillating behavior with bright, very asymmetric episodes followed by a darker, more symmetric episodes. The
period for the oscillation is about 4 minutes, i.e. below that associated with the acoustic
cutoff frequency (about 5.3 mHz). This may be caused by the presence of inclined magnetic fields can lower effectively the acoustic cutoff frequency \citep{Bel+Leroy1977}. The
time evolution of the 8542 \AA\ Stokes $I$ has a diagonal structure moving from blue
to red. This indicates the presence of propagating
compressible waves \citep{Carlsson+Stein1997}. The bright part, which corresponds to a large self-reversal, is clearly shifted towards the blue. This is seen in the 8498
\AA\ line profiles as well, although these profiles tend to be more
flat-bottomed. In general, the self-reversals and over all variation is
larger in the blue wing than in the red. This is true for all
slit positions which exhibit strong time-dependent behavior.

\begin{figure*}
\epsscale{0.8} \plotone{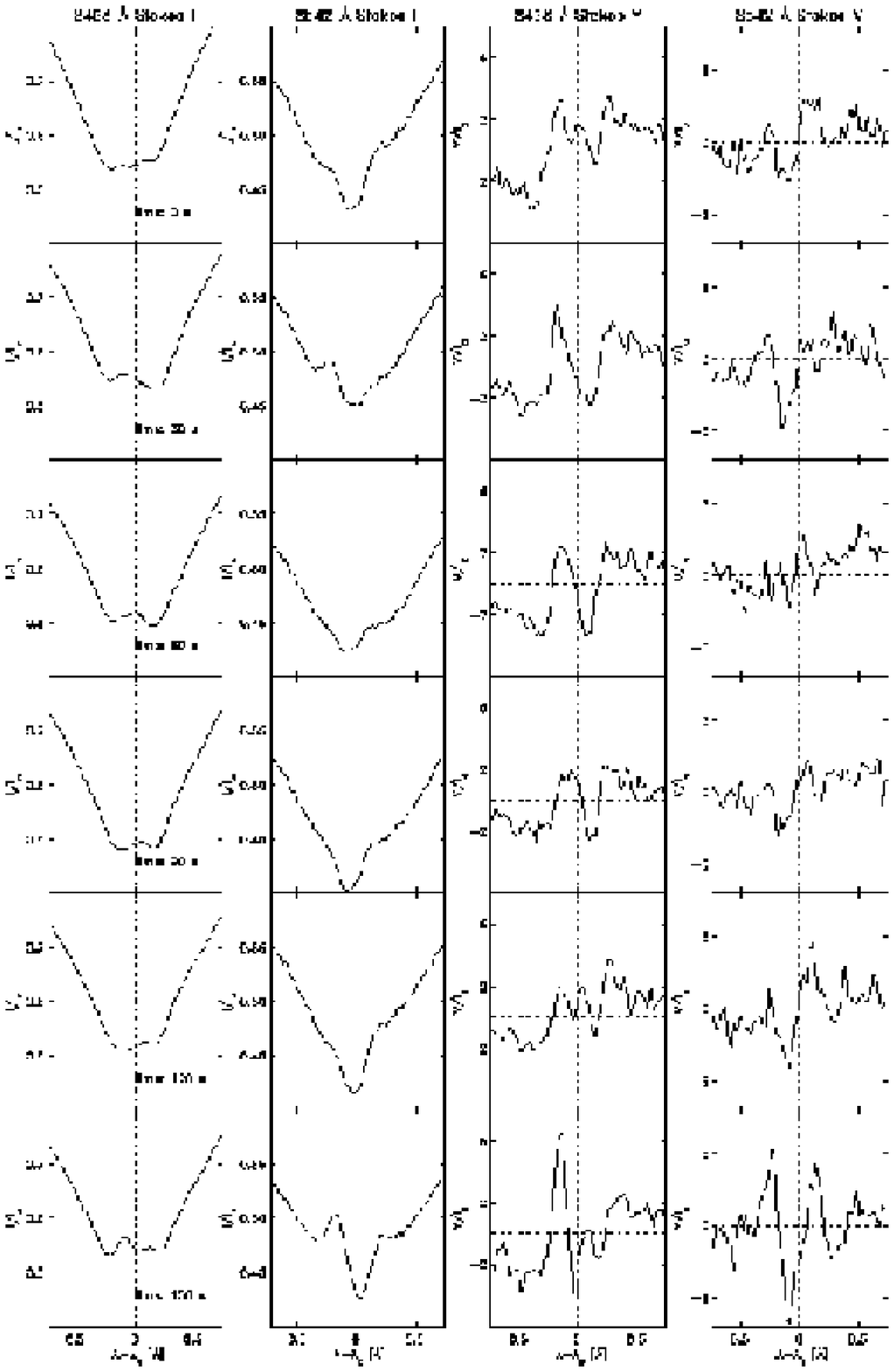}
\caption{Time evolution of individual Stokes $I$ and $V$ profiles in an intermediate flux pixel.  Location of the pixel is marked with a diamond in
Figure \ref{fig:magnetogram}.
\label{fig:ts40}
}
\end{figure*}

The Stokes $V$ image of the 8498 \AA\ line also shows strong diagonal
structures that coincide in time with the dark phases of Stokes
$I$. Inspection of individual profiles (Fig. \ref{fig:ts40}) reveals a pattern of multiple
lobes in the Stokes $V$ profiles. These lobes are on the blue side of
the line core and their amplitudes and positions vary periodically in
time resulting in the diagonal structure seen in the image. The lobes can
be identified with the emission features seen in the Stokes $I$
profiles. The 8542 \AA\ line Stokes $V$ image shows a pattern of a
multi-lobed profiles whose amplitudes vary strongly in time. The large
Stokes $V$ amplitude phase coincides with the bright, very asymmetric
phase seen in the intensity profiles. The red
wings always exhibit less structure and variation than the blue wings.


\subsubsection{Strong network}

\begin{figure*}
\epsscale{1} \plotone{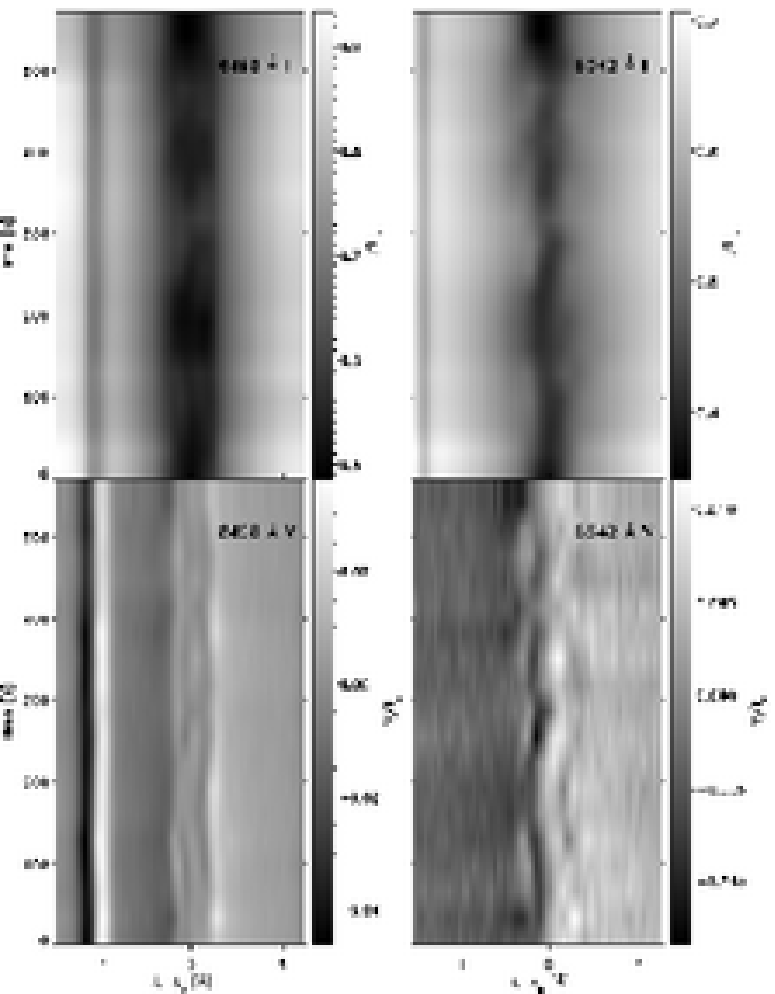}
\caption{Time evolution of Stokes $I$ and $V$ in a network pixel. Location of the pixel is marked with a triangle in
Figure \ref{fig:magnetogram}.
\label{fig:image79}
}
\end{figure*}

\begin{figure*}
\epsscale{0.8} \plotone{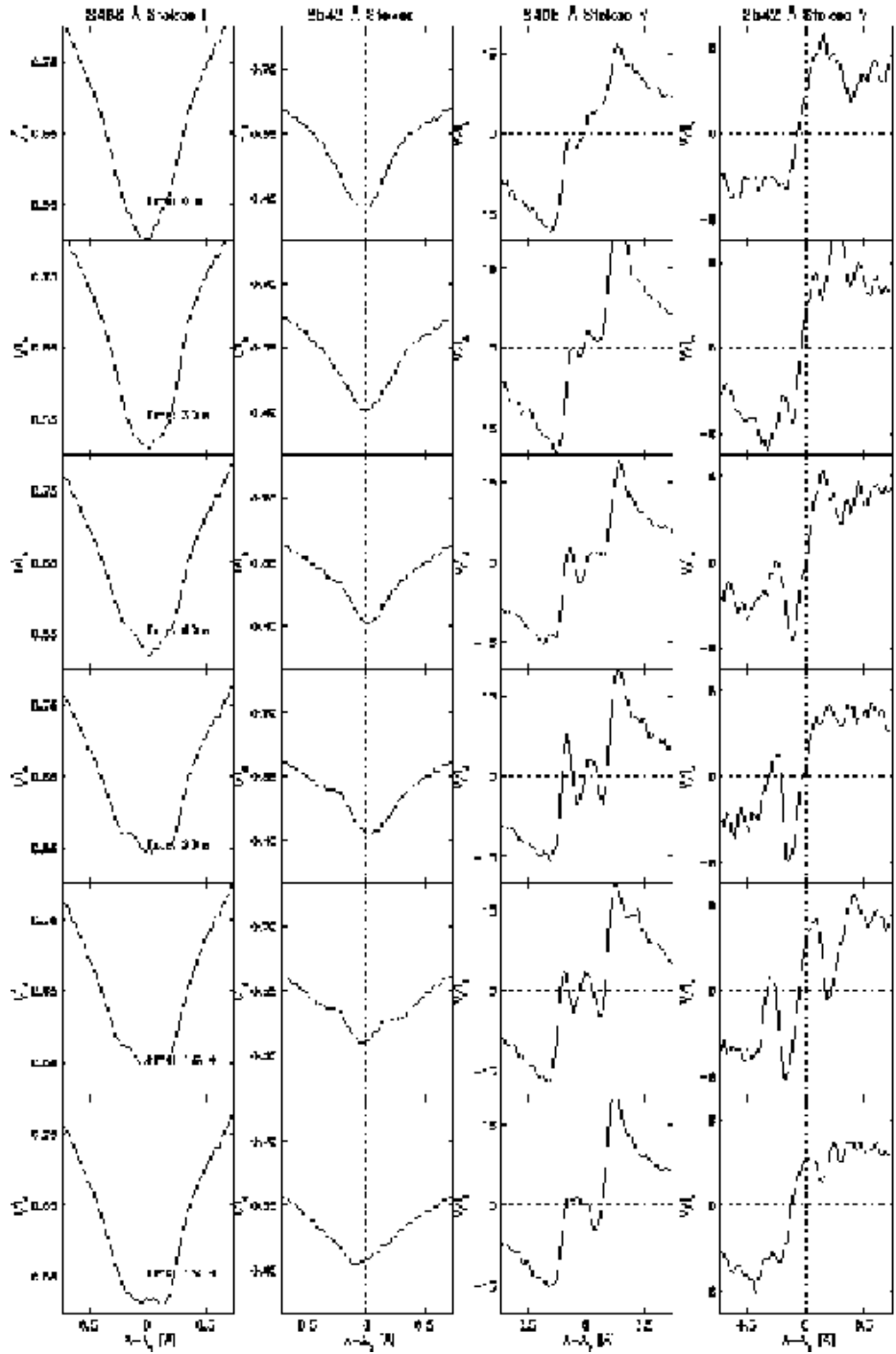}
\caption{Time evolution of individual Stokes $I$ and $V$ profiles in a network pixel.  Location of the pixel is marked with a triangle in
Figure \ref{fig:magnetogram}.
\label{fig:ts79}
}
\end{figure*}

Stokes $I$ and $V$ profiles seen in the strong network regions (Figs. \ref{fig:image79} and
\ref{fig:ts79}) would appear at first glance to be a mixture
of the less dynamic internetwork and the highly dynamic intermediate
flux region. The Stokes $I$ profiles exhibit the same pattern of
bright (more asymmetric) and dark (less asymmetric) phases as seen in
the intermediate flux region. The difference between the two phases is
however not as large: the amplitude of the self-reversals, especially
in the 8542 \AA\ intensity profiles, is much smaller than in the
intermediate flux case.

The Stokes $V$ images resemble those of the intermediate flux
region: some diagonal structures are seen, but they are weaker. The
8542 \AA\ line Stokes $V$ profiles have a time varying amplitude but
the profiles are not as asymmetric and they are not necessarily
multi-lobed. The difference between the time-dependent behavior of the
red and blue lobes of the profile, i.e. the red lobe varies less in
time, is even more clear here than in the intermediate flux region.

\subsection{Statistics}
\label{sec:stat}

\begin{figure*}
\epsscale{0.7} \plotone{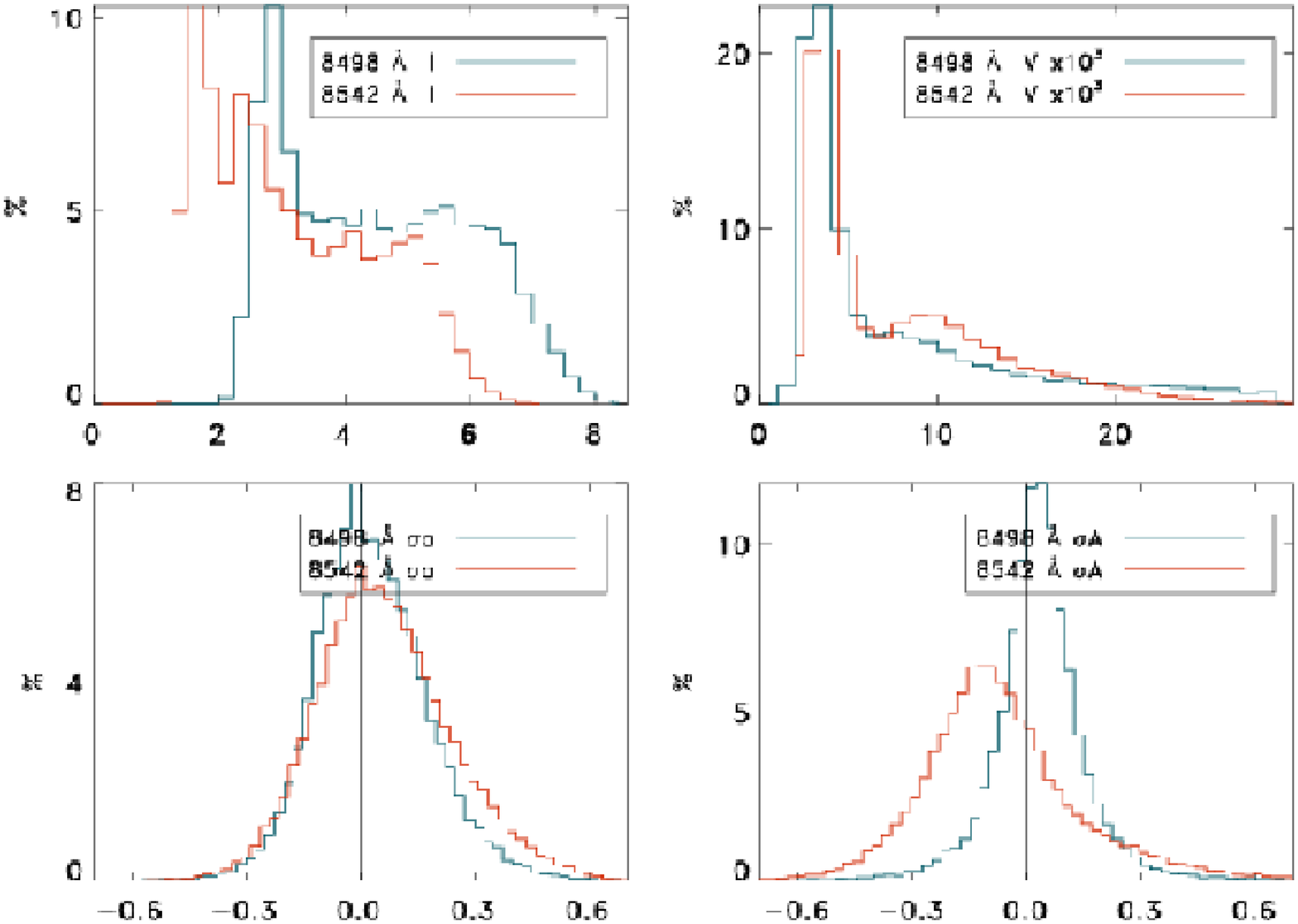}
\caption{Histograms of Stokes $I$ and $V$ amplitudes, and Stokes $V$ amplitude and area asymmetries of the map and time-series.  
\label{fig:hgram}
}
\end{figure*}

Histograms of the Stokes $I$ amplitude integrated over 250 m\AA\ around the line core for the two Ca
lines are shown in the top-left panel of Figure ~\ref{fig:hgram}. These histograms include both the map and time-series profiles. Because there are almost five times as many profiles in the time-series as there are in the map, the histograms are dominated by the time-series profiles. Both
lines exhibit a wide range of values. Except for the peaks at low intensities, the histograms are fairly flat. The
darkest (i.e. lowest core intensity or most absorption) amplitudes, are associated with
the internetwork, and the brightest with the network.
 
Histograms of the Stokes $V$ amplitudes (top right panel of
Fig. \ref{fig:hgram}) peak at the same value in both lines, 0.003
$I_c$, but the 8498 \AA\ histogram tail decays more slowly. Since the 8498
\AA\ line is formed slightly lower of the two and the lines are
roughly equally sensitive to magnetic fields (effective Land\'e g
factors are 1.07 and 1.10 for the 8498 \AA\ and the 8542 \AA\ lines,
respectively), it is not surprising that the 8498 \AA\ histogram has
the longer tail.

Both lines' Stokes $V$ amplitude asymmetry histograms (bottom left
panel of Fig. \ref{fig:hgram}) have very similar shapes and similar
widths. There are more positive asymmetries in both lines: 56\% in
8498 \AA\ and 64 \% in 8542 \AA\ (Table \ref{tab:asym}). The mean
amplitude asymmetries are also positive, and the 8542 \AA\ mean
asymmetry is two times larger. There are more negative amplitude
asymmetries in the 8542 \AA\ map than in the time series. Non-zero
amplitude asymmetries indicate at least one of two things: the spatial
pixels consist in most cases of at least two atmospheric components
that are shifted relative to one another or that there are velocity
and/or magnetic field gradients present in the atmosphere.

The area asymmetry histograms (bottom right panel of
Fig. \ref{fig:hgram}) of the two calcium lines repeat the pattern
already seen in the cluster profiles: the 8542 \AA\ histograms
is centered around a negative value and the 8498 \AA\ is centered at
roughly zero, though the mean is slightly positive. The 8542 \AA\
histogram is significantly wider than the 8498 \AA\ histogram. A
multi-component atmosphere alone cannot produce area asymmetries, so
the existence of non-zero area asymmetries indicates the presence of
velocity and possibly magnetic gradients in the atmosphere.

In the 8542 \AA\ line 66 \% of the profiles have negative area
asymmetries whereas in the 8498 \AA\ line the majority of the
profiles, 64 \%, have positive area asymmetries (Table
\ref{tab:asym}). To better understand why the area asymmetry
histograms of the lines are so different, we need to look at the
components of the area asymmetry separately i.e. the sign of the blue
lobe and the total area of the Stokes $V$ profile. One possible cause
for the difference in the histograms might be that the distribution of
signs of the blue lobe is different in the two lines. Closer
inspection reveals that this is not the explanation. The vast majority
of both lines, over 80 \%, have a negative sign. (Here the sign is
defined to be the sign of the local maximum or minimum amplitude of the blue lobe). A
second possible explanation is that the 
$\int_{\lambda_0}^{\lambda_1} V(\lambda)d\lambda$ is different in the two lines. This is found to be the
case. The 8542 \AA\ line has more profiles with a positive area and
the 8498 \AA\ has slightly more profiles with a negative area. (Note
that the sign of the area asymmetry is the product of the sign of the
blue lobe and the area; eq. \ref{eqn:area}.) The area of the Stokes $V$ profile is
strongly affected by the emission features. These features, and their
amplitudes, are related to the self-reversals seen in the Stokes $I$
profiles. The self-reversals are stronger on the blue side of the line
core than on the red. In general, the blue lobes of the Stokes $V$ profiles have negative
amplitudes and the effect of the emission features is then to reduce
the amplitude, and in some cases, make it positive and this way reduce
the overall negative area. 

The effect of the emission features on the amplitude asymmetries is
not as large because the amplitude will be affected only if the
emission feature is located at the same wavelength as the maximum
absolute amplitude. Also if the profile has a
wide blue lobe, i.e., the wings contribute significantly, a local
reduction in peak amplitude is counterbalanced by a comparable
signal in the other parts of the blue lobe. The resulting profile will
have nearly the same amplitude in the blue lobe as before, but the area will be reduced leading to a smaller, or even negative, area asymmetry. Since
the self-reversals are larger in the 8542 \AA\ line, this scenario is
more likely to apply to it than the 8498 \AA\ line. 

Both lines' area and amplitude asymmetries are found
to be inversely proportional to the Stokes $V$ amplitudes. The
scatter, especially in the 8542 \AA\ line, is fairly large.

PCA also allows us to ensure that the determination of Stokes $V$
asymmetries is not dominated by noise. Reconstructing the profiles
using only the 11 first eigenvectors (i.e., essentially noise-free
profiles) and then computing the asymmetries reproduces the Stokes $V$
amplitude and asymmetry histograms.  To test if the negative histogram
peak in the 8542 \AA\ line is an artifact caused by data reduction, we
computed area asymmetries for the datasets, but after first removing
the fringe pattern caused by the optics. This did not alter the area
asymmetry histogram. Another artifact that could cause the offset is
an incorrect subtraction of the tilt caused by the detector in the
continuum intensity. To remove the offset in the histograms by means
of changing the tilt causes a clearly visible lopsidedness in the
Stokes I profiles. Lastly, to make sure that the choice of the integration
range is not the cause of the offset, we used a constant bandwidth for
area asymmetries and it also reproduces the 8542 \AA\ area histogram
offset. (Besides these issues, there are no other obvious artifacts
that would cause the offset.) We therefore conclude that the offset is
not caused by the fringing or incorrect subtraction of the tilt in the continuum intensity.

\section{Comparison of observations with a high-$\beta$ simulation}
\label{sec:highb}

In P06 we synthesized Stokes profiles for
the Ca IR triplet lines in the high-$\beta$ regime. This was done by
combining a radiation hydrodynamic code (see for example
\citealt{Carlsson+Stein1997}) with a weak magnetic field and using a
nLTE Stokes inversion and synthesis code
\citep{Socas-Navarro+others2000} to produce, based on
snapshots of the simulation, a time series of the lines' Stokes vectors. The simulation is driven by a photospheric velocity piston and
its dynamics are dominated by upward propagating acoustic
waves in a simple magnetic field topology.
The simulation shows that the radiative transfer is very
similar in all the Ca IR triplet lines. The differences between the
line behaviors in the simulation are mainly due to the lines having
slightly different formation heights and thus experiencing a
difference in the amplitudes of the shocking waves: the higher the
line is formed, the larger the amplitude of the passing wave is.

In the
simulation there is no feedback from the magnetic fields on the
dynamics and the waves are purely acoustic. The observations have
limited spatial and temporal resolutions whereas the simulation is much better resolved.

\subsection{Comparison of time dependent behavior}

\begin{figure*}
\epsscale{1} \plotone{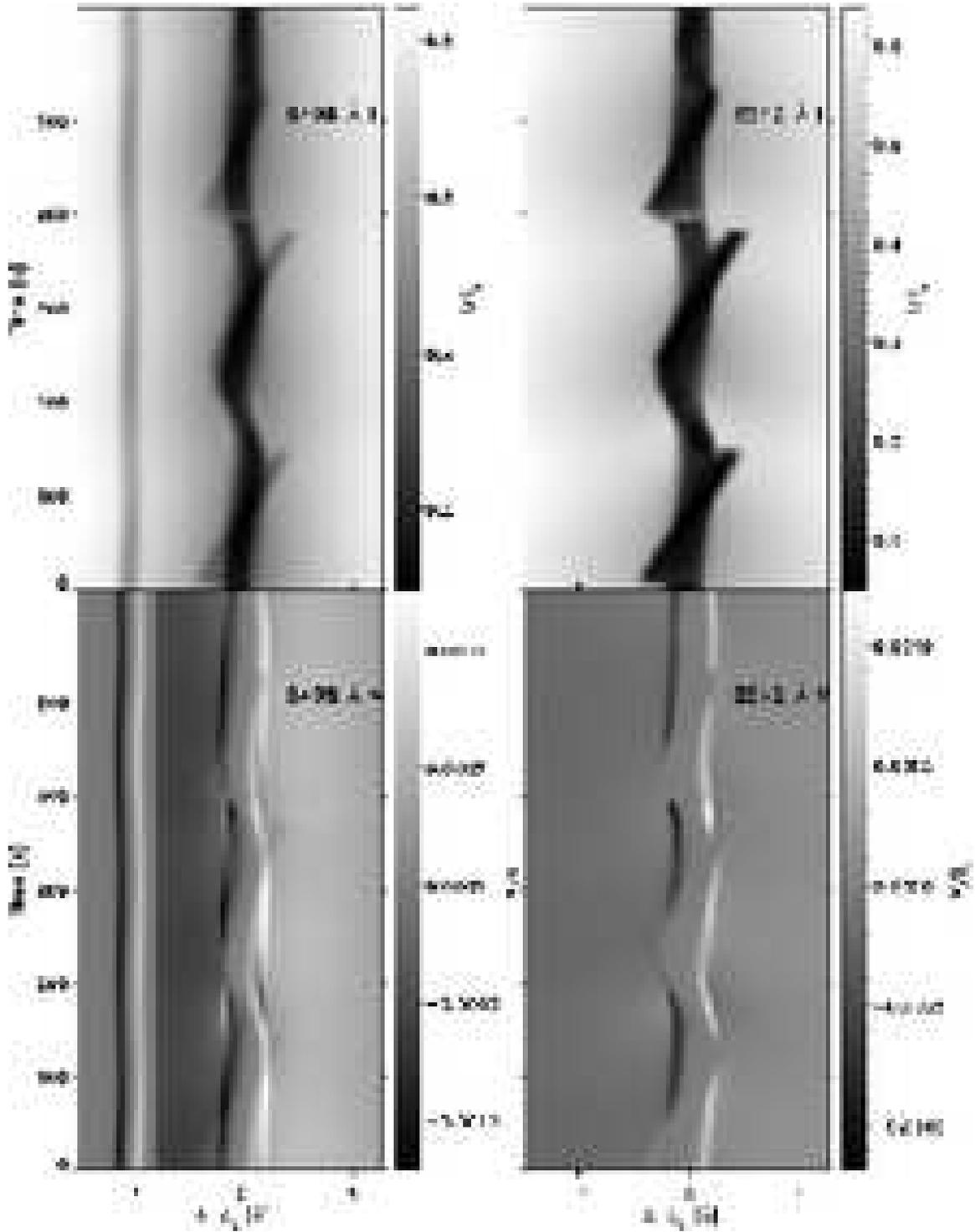}
\caption{Time evolution of Stokes $I$ and $V$ profiles in the high-$\beta$ simulation (P06). The Stokes V signal in wavelength range -1.2 to -0.6 \AA\ in the 8498 \AA\ image is scaled down with factor 7.5 in order to display both the \ion{Ca}{2} 8498 \AA\ and \ion{Fe}{1} 8497 \AA\ lines in the same panel. 
\label{fig:simts}
}
\end{figure*}

As the acoustic waves in the simulation propagate upwards and
eventually form shocks, a time-varying pattern of disappearing and
reappearing Stokes $V$ lobes is seen (Fig.~\ref{fig:simts}). The
pattern is strongest in the highest forming line, i.e. 8542 \AA. Wave
propagation is also seen in the Stokes $I$ profiles. There are no
large self reversals or brightenings, instead the position of the line
minimum changes periodically and forms a
saw-tooth like pattern where the red shift takes more time than the
blue shift phase.

If we first compare the simulated profiles to the internetwork
observations (Fig. \ref{fig:image130}), we see that the strong
signatures of shocks seen in the simulation are not present in the observations. In the simulation the Ca IR triplet is formed in a region where the
waves are just beginning to shock. If the formation height of the
lines or the shocks in the simulation is off, compared to the real
Sun, by a small amount, even 50 km, the lines' temporal evolution may
look very different. Another possible explanation to why we see no strong indications of
shocks is the temporal and/or spatial resolution: there may be several components
oscillating out of phase relative to one another in a given resolution
element. However, the photospheric velocities are very similar in the
internetwork and network, but the network profiles show strong
self-reversals. This suggests that spatial and temporal resolution
alone cannot explain the lack of strong signatures of shocks in the
internetwork.

Observations of the quiet Sun show varying
degrees of oscillatory power (compare for example \citet{Lites+Rutten+Kalkofen1993a} [\ion{Ca}{2} H and K] or UV data of
\citealt{Judge+Carlsson+Stein2003}, \citealt{McIntosh+Judge2001} and
\citealt{Wikstol+others2000}). This variation may be related to the
local magnetic topology, especially to the possible existence of a
magnetic canopy \citep{McIntosh+Fleck+Judge2003,Vecchio+others2006}. The
region observed here was less oscillatory than average but still not
exceptionally quiet.
 
Both the simulated profiles and observed network profiles (Fig.~\ref{fig:image79}) show time varying patterns
where the Stokes $I$ and $V$ amplitudes change periodically. In the
simulation the wave propagation manifests itself in the Stokes $I$
profiles most clearly as a shift of the line core and the saw-tooth shape of the time series. In the observations, waves cause the
lines' periodically varying self-reversals that result in
alternating bright and dark phases. There are indications of
diagonal structures in the observed Stokes $I$ images, but they are
not nearly as clear as in the simulation. In the simulation the
upward propagating waves cause the blue and red lobes of the Stokes
$V$ profiles to disappear alternately. In contrast, the observed time varying
pattern in Stokes $V$ looks more complicated: there is much more structure in the observed profiles, especially in the line cores, than in the simulation. This is related to the simulated profiles not exhibiting
strong self-reversals as seen in the observations. 

In the simulation, because of radiative cooling and expansion of the
falling material, the down flows are in general cooler than the
up flows. In the synthesized profiles this manifests itself by the red
wings of the Stokes $I$ profiles showing less variations, though the difference with the blue wing is quite small. Similar
behavior is also seen in the observations: the self-reversals are in
general larger in the blue wing of the Stokes $I$ profiles and the red
lobes of the Stokes $V$ profiles show clearly less variation.

\subsection{Comparison of statistics  and Stokes $V$ morphologies}

In the simulation the magnetic field decays exponentially with height
and therefore the \ion{Ca}{2} Stokes $V$ amplitudes are significantly
lower than the \ion{Fe}{1} 8497 \AA\ amplitude. In the observations the Ca and Fe line Stokes $V$ profiles have roughly the same amplitudes. This may be explained by the field decaying much slower with height in the observations, or by the filling factor in the observations being smaller in the photosphere than in the chromosphere.  

Both \ion{Ca}{2} lines' observed  Stokes $V$ profiles have a
significant amount of signal in the wings. In the simulations only the
8498 \AA\ line Stokes $V$ has extended wings with large amplitudes
(Fig. 4 in P06). The amount of signal in the
wings depends on the atmospheric magnetic field gradient. If there is
no gradient the wings of all three Ca lines have very little signal. Whereas a
model atmosphere with a constant field gradient produces profiles
where all lines, 8498 \AA\ the most, have some signal in the line
wings and an exponential field produces profiles with the largest
wings. Depending on where the gradient is located and how strong the
field is, the Ca lines may or may not have similar Stokes $V$
profiles. Based on the profile shapes and relative amplitudes, it is
obvious that the magnetic topology in the observations is
different from the simulation.

\begin{figure*}
\epsscale{1} \plotone{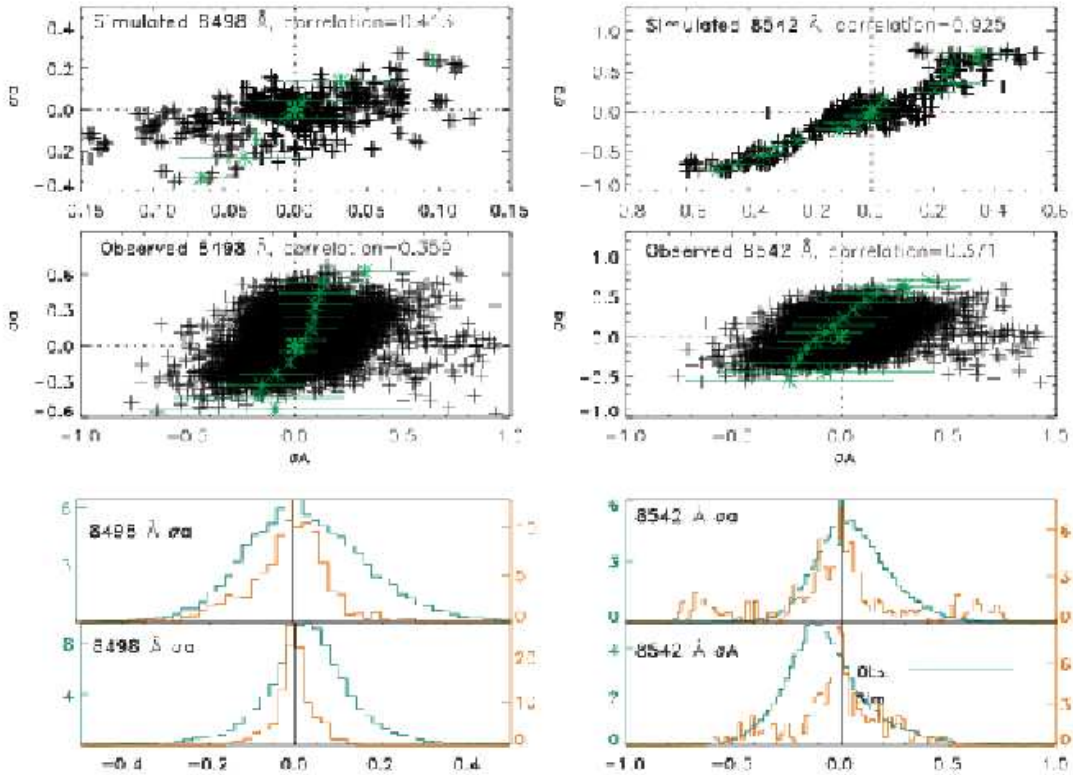}
\caption{Stokes $V$ asymmetries of the simulated and observed profiles. Upper 4 panels show the correlation of amplitude and area asymmetries in the simulated and observed Ca lines. The Pearson correlation coefficient for each case is given. The asterisk symbols show the mean for each 0.1 wide bin and the error bars show the standard deviation. The lower panels are histograms of observed and simulated amplitude and area asymmetries. 
\label{fig:simobs}
}
\end{figure*}

Formation of area and amplitude asymmetries in the simulation is
coupled. The correlation is especially strong in the 8542 \AA\ line (upper
row of Fig. \ref{fig:simobs}). In the 8498 \AA\ Stokes $V$
profiles the strong wings affect the asymmetries, and the
correlation is weaker. The observed area and amplitude
asymmetries of both lines show less correlation. This is at least partly because the
observed profiles have more complex shapes than in the simulation.

The lower panels in Figure \ref{fig:simobs} show the Stokes $V$
asymmetry histograms for the simulation. The observed histograms are
re-plotted to enable direct comparison. In the simulation both lines'
amplitude and asymmetry histograms are centered roughly around zero
(percentage-wise there are a couple of percent more negative than
positive asymmetries). This was not the case in the observations
where all the asymmetries, except the 8542 \AA\ area asymmetry, have
clearly more positive than negative values, i.e. the blue lobe is
larger in area/amplitude than the red lobe.

The observed 8498 \AA\ profiles are more
dynamic than the simulated ones. Consequently the observed 8498 \AA\
asymmetry histograms are clearly wider than the simulated.  Because there is very little signal in the simulated 8542 \AA\ Stokes $V$ profile wings, when an upward propagating wave
causes a Stokes $V$ lobe to disappear, there is no signal in the line
wing to contribute to the amplitude. This leads to the extreme
amplitude asymmetries in the simulations and in the
additional lobes at large values in the simulated 8542 \AA\ line area
asymmetry histogram. Since the observed profiles
have a significant amount of signal in the wings, the extreme
amplitude asymmetries are moderated, and no lobes at large values are seen in the histogram.

\section{Conclusions and Discussion}
\label{sec:end}

So far most spectropolarimetric studies using the \ion{Ca}{2} IR
triplet lines have focused on active regions ({\it e.g.,} \citealt*{Socas-Navarro+others2000b, LopezAriste+others2001, Socas-Navarro2005a,Uitenbroek+others2006}). The observations
presented here show that these lines are also promising candidates for
studying the magnetic chromosphere outside of active
regions. Interpreting the observations, however, is not straight
forward.

The main results of the analysis presented here are:

\begin{itemize}
\item Classification of Stokes $V$ profile shapes. 

Asymmetric line profiles are
very common and that the two lines, despite being formed fairly close in a geometrical sense,
often do not have similar shapes. Furthermore, the edges of the
network patches exhibit profile shapes different from those seen in
the center of the patches. The cluster analysis results, as expected, in a qualitative, not quantitative, description of the profile shapes.

\item Statistics of the line profiles.

The 8542 \AA\ area asymmetry is predominantly
 negative; while the 8498 \AA\ area asymmetry and the amplitude asymmetries are usually positive. 

\item Time dependent behavior. 

The enhanced network has very different dynamic behavior compared with
the internetwork. It is more dynamic and the oscillation period, as seen
in both Stokes $I$ and $V$, is greater than in the internetwork.

\item Comparison with high-$\beta$ simulation.

Oscillations are present in both the observations and the simulation. The
simulated profiles are more dynamic than the observed internetwork
profiles. The opposite is true for network profiles. In the simulation,
the formation of asymmetries is more tightly coupled than what is seen in the observations. Except for the 8542 \AA\ amplitude asymmetry the
observed profiles show a wider range of asymmetries. And lastly, the
peculiar negative area asymmetries seen in the observed 8542 \AA\ line
and the tendency of the other asymmetries to be positive are not reproduced 
by the simulation.

\end{itemize}

The tendency of large Stokes $V$ asymmetries to decrease with an
increasing signal amplitude has also been observed in photospheric lines
\citep{Grossmann-Doerth+Keller+Schussler1996}. In the photosphere a magnetic canopy is one possible explanation: the canopy gives rise to
asymmetries in the lines, and as a flux tube diameter increases, the
relative contribution from the canopy to the Stokes $V$ signal
decreases. In the photosphere the scatter in an amplitude vs. asymmetry plot is significantly larger in
the area than in the amplitude. No large difference is
seen in the area and amplitude asymmetry scatters of the
\ion{Ca}{2} lines. 

In the quiet Sun photosphere, more positive than negative Stokes $V$
asymmetries are found \citep{Grossmann-Doerth+Keller+Schussler1996}. In
contrast with 8498 \AA\ line (where there is no large difference in the mean area and amplitude asymmetries) the
photospheric mean area asymmetries are significantly smaller (4 \%\ in the
\ion{Fe}{1} 6302 \AA\ line) than the mean amplitude asymmetries (15
\%\ in the \ion{Fe}{1} 6302 \AA\ line). The photospheric asymmetries
are often attributed to multiple atmospheric components within a
resolution element. In the chromosphere, however, gradients have to
play a dominant role since the formation of area asymmetries require
them. Another piece of evidence of the importance of gradients in the chromosphere is that Milne-Eddington inversions, which include the Paschen-Back effect of the \ion{He}{1} 10830 \AA\ triplet, are not able to reproduce the observed area asymmetries \citep{Sasso+Lagg+Solanki2006}.

\citet{Khomenko+others2005} used a 3-dimensional magnetoconvection model to synthesize photospheric magnetically sensitive lines in the visible and IR. There are more positive than negative Stokes $V$ asymmetries in their synthetic profiles. 
They found that reducing the spatial resolution increases the number of
irregular stokes V profiles (though the number of strongly asymmetric profiles decreases). They conclude that the asymmetries reflect
more inhomogeneities in the horizontal direction than in the
vertical. In the chromosphere large velocity gradients are more common and variation in the vertical direction are likely to be more important than variation in the horizontal direction. When these two factors are combined with the observed area asymmetries, one concludes that the
chromospheric asymmetries mainly reflect the line-of-sight inhomogeneities, and not variations in the horizontal direction. Despite the apparent similarities between the photospheric and
chromospheric Stokes $V$ profiles, the underlying mechanism causing
the asymmetries does not appear to be the same. Drawing parallels between the
chromosphere and photosphere is problematic since the two regions exist in
very different physical regimes.

The discrepancy between the Stokes $V$ asymmetry histograms of
 the
observations and the simulation may be related to the self-reversals. The
simulated profiles exhibit only small self-reversals. The
observations show large self-reversals in the Stokes $I$ profiles and
accompanying emission features in the Stokes $V$ profiles. These features
are stronger on the blue side of the line cores. Another effect that contributes to the
imbalance is that that the down flow phase lasts longer. Our
observations, especially with a 5 second exposure time, sample more
profiles with red-shifts and positive asymmetries (since there will be
more emission on the blue side). However, inspection of Fig. \ref{fig:simts} shows the same to be true of the simulations. If this is the case, why are there not more
positive than negative asymmetries in the simulation as well?

The sample of these observations is limited because the majority of the
profiles are drawn from the same three slit positions which sample the same local
magnetic field configuration. It would not be surprising if histograms
made of profiles from a variety of quiet-Sun magnetic field topologies
would have somewhat different shapes. The complexity of the observed
profiles makes the interpretation of the area and amplitude asymmetries
difficult. Because of multiple lobes and the strong signal in the line
wings, the asymmetries are not necessarily good proxies for the
overall complexity of the Stokes $V$ profiles. This is especially true
if the two asymmetries are viewed separately.

It is a well known result that the network intensity oscillations have a longer
period than the internetwork
(e.g. \citealt{Orrall1966}, \citealt{Lites+Rutten+Kalkofen1993a},
\citealt{Banerjee+others2001}). This has also been observed before in
the \ion{Ca}{2} IR lines \citep{Deubner+Fleck1990}. Why do the intermediate
flux regions in our observations appear to be more dynamic than the stronger flux regions? It may be related to a more complex magnetic
topology at the edges of the network patches. The observations show no signal above
the noise in Stokes $Q$ and $U$, so we cannot draw any conclusions of
possible horizontal fields. Any signal would be affected by atomic polarization \citep{MansoSainz+TrujilloBueno2003} making the interpretation exceedingly complex. The filling
factor in the network is not likely to be very large, and is likely smaller at the edges than in the center of the network patch. Inversions by
\citet{BellotRubio+others2000} of average Stokes profiles in a plage
region gave a filling factor of 0.5 a $z=0$ km. The filling factor in the
photospheric network can safely be assumed to be lower than this. In fact, in recent
inversions by \citet{Cerdena+others2006}, which included a small patch
of network, the photospheric filling factor in the patch center was as small as 0.1. The network magnetic fields must expand
with height and consequently  the chromospheric filling factor
must exceed photospheric values. Results of comparing photospheric and
chromospheric magnetograms, however, \citet{Zhang+Zhang2000} suggest that the sizes of the network
magnetic elements are not very different at the two heights
. The chromospheric magnetograms in the comparison are 
based on the H$\beta$ line. Its interpretation is complicated by the magnetically sensitive blends close to the line core, and the line may suffer from same problems as the H$\alpha$ line when used as a proxy for chromospheric magnetic fields, namely that the photospheric contribution to the polarization signal is not insignificant \citep{Socas-Navarro+Uitenbrock2004}. Lastly, the size of network patches is not directly linked with the filling factor. We see some expansion of the network with height in the magnetograms of the map (Fig. \ref{fig:magnetogram}), especially when comparing the \ion{Ca}{2} 8498 \AA\ and 8542 \AA\ magnetograms. But since the magnetograms were constructed by using the weak field formula, and the network fields have gradients and are not necessarily weak, the magnetograms are not accurate. Also the choice of color scaling of the images affects the comparison. However, the apparent expansion is not necessarily an artifact, since expansion of network seen in magnetograms has also been reported by \citet{Giovanelli1980}.

Obviously we need to understand better the topology of the network
magnetic fields. To do this we plan to perform nLTE inversions of these
data in the near future. The inversions will help further in understanding the formation dynamics of the \ion{Ca}{2} IR
lines in the quiet Sun, and hopefully reveal how the
underlying atmosphere differs from that used in the
simulation. An important question to answer is why the two Ca lines
behave as differently as they do. Having a
time series taken during good seeing would be helpful. Also in order to expand the analysis to
internetwork regions, better spatial resolution is required. Another
interesting question is how much variation there is in dynamics in different
internetwork regions, and how well the differences can be
explained in terms of the surrounding magnetic fields as has been suggested by \citet{Vecchio+others2006} based on imaging data of \ion{Ca}{2} 8542 \AA\ Stokes $I$. To fully investigate this in detail high quality data of the full Stokes vector are needed. 

\clearpage

\begin{deluxetable}{ccccccc}
\tablewidth{0pt}
\tablecaption{Observed Stokes $V$ asymmetries}
\tablehead{\colhead{} & \colhead{8498 \AA } & \colhead{8498 \AA} & \colhead{8498 \AA} & \colhead{8542 \AA} & \colhead{8542 \AA} & \colhead{8542 \AA} \\ 
\colhead{} & \colhead{$<0$ (\%)} & \colhead{$>0$ (\%)} & \colhead{mean (\%)} & \colhead{$<0$ (\%)} & \colhead{$>0$ (\%)}  & \colhead{mean (\%)}}

\startdata
\hline
$\sigma$a & 43.2 & 55.7 & 3.1 & 36.6 & 61.4 & 6.3 \\
$\sigma$A & 35.5 & 64.5 & 3.3 & 69.7 & 30.3 & -6.8\\

\enddata
\label{tab:asym}
\tablecomments{Percentages of observed \ion{Ca}{2} 8498 \AA\ and 8542 \AA\ Stokes $V$ amplitude and area asymmetries with negative (i.e. red lobe larger) and positive (i.e. blue lobe larger) signs.
}


\end{deluxetable}

\clearpage

\acknowledgments{Thanks to Doug Gilliam, Joe Elrod and Mike Bradford for all their invaluable help during the observing run.

\begin{thebibliography}{39}
\expandafter\ifx\csname natexlab\endcsname\relax\def\natexlab#1{#1}\fi


\bibitem[{{Andretta} \& {Jones}(1997)}]{Andretta+Jones1997}
{Andretta}, V. \& {Jones}, H.~P. 1997, \apj, 489, 375

\bibitem[{{Banerjee} {et~al.}(2001){Banerjee}, {O'Shea}, {Doyle}, \&
  {Goossens}}]{Banerjee+others2001}
{Banerjee}, D., {O'Shea}, E., {Doyle}, J.~G., \& {Goossens}, M. 2001, \aap,
  371, 1137

\bibitem[{{Bel} \& {Leroy}(1977)}]{Bel+Leroy1977} {Bel},N. \& {Leroy}, B. 1977, \aap, 55, 239


\bibitem[{{Bellot Rubio}(2006)}]{BellotRubio2006}
{Bellot Rubio}, L.~R. 2006, ArXiv Astrophysics e-prints

\bibitem[{{Bellot Rubio} {et~al.}(2000){Bellot Rubio}, {Ruiz Cobo}, \&
  {Collados}}]{BellotRubio+others2000}
{Bellot Rubio}, L.~R., {Ruiz Cobo}, B., \& {Collados}, M. 2000, \apj, 535, 489

\bibitem[{{Carlsson} \& {Stein}(1997)}]{Carlsson+Stein1997}
{Carlsson}, M. \& {Stein}, R.~F. 1997, ApJ, 481, 500

\bibitem[{Deubner \& Fleck(1990)}]{Deubner+Fleck1990}
Deubner, F.-L. \& Fleck, B. 1990, \aap, 228, 506

\bibitem[{{Dom{\'{\i}}nguez Cerde{\~n}a} {et~al.}(2006){Dom{\'{\i}}nguez
  Cerde{\~n}a}, {Almeida}, \& {Kneer}}]{Cerdena+others2006}
{Dom{\'{\i}}nguez Cerde{\~n}a}, I., {Almeida}, J.~S., \& {Kneer}, F. 2006,
  \apj, 646, 1421

\bibitem[{Giovanelli(1980)}]{Giovanelli1980}
Giovanelli, R.~G. 1980, \solphys, 68, 49

\bibitem[{{Grossmann-Doerth} {et~al.}(1996){Grossmann-Doerth}, {Keller}, \&
  {Schuessler}}]{Grossmann-Doerth+Keller+Schussler1996}
{Grossmann-Doerth}, U., {Keller}, C.~U., \& {Schuessler}, M. 1996, \aap, 315,
  610

\bibitem[{{Judge} {et~al.}(2003){Judge}, {Carlsson}, \&
  {Stein}}]{Judge+Carlsson+Stein2003}
{Judge}, P.~G., {Carlsson}, M., \& {Stein}, R.~F. 2003, \apj, 597, 1158

\bibitem[{{Keller} \& {The Solis Team}(2001)}]{Keller2001}
{Keller}, C.~U. \& {The Solis Team}. 2001, in ASP Conf. Ser. 236: Advanced
  Solar Polarimetry -- Theory, Observation, and Instrumentation, ed.
  M.~{Sigwarth}, 16--+

\bibitem[{{Khomenko} {et~al.}(2003){Khomenko}, {Collados}, {Solanki}, {Lagg},
  \& {Trujillo Bueno}}]{Khomenko+others2003}
{Khomenko}, E.~V., {Collados}, M., {Solanki}, S.~K., {Lagg}, A., \& {Trujillo
  Bueno}, J. 2003, \aap, 408, 1115

\bibitem[{Khomenko} {et~al.}(2005)]{Khomenko+others2005}{Khomenko}, E.~V., {Shelyag}, S., {Solanki}, S.~K. \& {V{\"o}gler}, A. 2005, \aap, 442, 1059

\bibitem[{{Lagg}(2005)}]{Lagg2005}
{Lagg}, A. 2005, in ESA SP-596: Chromospheric and Coronal Magnetic Fields, ed.
  D.~E. {Innes}, A.~{Lagg}, \& S.~A. {Solanki}

\bibitem[{{Landi Degl'Innocenti}(1992)}]{LandiDeglInnocenti1992}{Landi Degl'Innocenti},E. 1992, in Solar Observations: Techniques and Interpretation, First Canary Islands Winter School of Astrophysics, ed. F. {Sanchez}, M. {Collados} \& M. {Vazquez}

\bibitem[{{Lites} {et~al.}(1982){Lites}, {Chipman}, \&
  {White}}]{Lites+Chipman+White1982}
{Lites}, B.~W., {Chipman}, E.~G., \& {White}, O.~R. 1982, \apj, 253, 367

\bibitem[{Lites {et~al.}(1993)Lites, Rutten, \&
  Kalkofen}]{Lites+Rutten+Kalkofen1993a}
Lites, B.~W., Rutten, R.~J., \& Kalkofen, W. 1993, \apj, 414, 345

\bibitem[{{L{\'o}pez Ariste} {et~al.}(2001){L{\'o}pez Ariste}, {Socas-Navarro},
  \& {Molodij}}]{LopezAriste+others2001}
{L{\'o}pez Ariste}, A., {Socas-Navarro}, H., \& {Molodij}, G. 2001, \apj, 552,
  871

\bibitem[{{MacQueen}(1967)}]{MacQueen1967}
{MacQueen}, J. 1967, in Proceedings Fifth Berkeley Symposium on Math. Stat. and
  Prob., ed. L.~M. {LeCam} \& J.~{Neyman}, 281--+

\bibitem[{{Manso Sainz} \& {Trujillo
  Bueno}(2003)}]{MansoSainz+TrujilloBueno2003}
{Manso Sainz}, R. \& {Trujillo Bueno}, J. 2003, Physical Review Letters, 91,
  111102

\bibitem[{{Mart{\'{\i}}nez Pillet}(1997)}]{MartinezPillet+others1997}
{Mart{\'{\i}}nez Pillet}, V., L. B.~W. . S.~A. 1997, \apj, 474, 810

\bibitem[{{McIntosh} {et~al.}(2003){McIntosh}, {Fleck}, \&
  {Judge}}]{McIntosh+Fleck+Judge2003}
{McIntosh}, S.~W., {Fleck}, B., \& {Judge}, P.~G. 2003, \aap, 405, 769

\bibitem[{{McIntosh} \& {Judge}(2001)}]{McIntosh+Judge2001}
{McIntosh}, S.~W. \& {Judge}, P.~G. 2001, \apj, 561, 420

\bibitem[{Neckel \& Labs(1984)}]{Neckel+Labs1984a}
Neckel, H. \& Labs, D. 1984, \solphys, 90, 205

\bibitem[{{Noyes}(1967)}]{Noyes1967}
{Noyes}, R.~W. 1967, in IAU Symp. 28: Aerodynamic Phenomena in Stellar
  Atmospheres, ed. R.~N. {Thomas}, 293--+

\bibitem[{Orrall(1966)}]{Orrall1966}
Orrall, F.~Q. 1966, \apj, 143, 917

\bibitem[{{Paletou} \& {Molodij}(2001)}]{Paletou+Molodij2001}
{Paletou}, F. \& {Molodij}, G. 2001, in ASP Conf. Ser. 236: Advanced Solar
  Polarimetry -- Theory, Observation, and Instrumentation, ed. M.~{Sigwarth},
  9--+

\bibitem[{{Pietarila} {et~al.}(2006){Pietarila}, {Socas-Navarro}, {Bogdan},
  {Carlsson}, \& {Stein}}]{Pietarila+others2006}
{Pietarila}, A., {Socas-Navarro}, H., {Bogdan}, T., {Carlsson}, M., \& {Stein},
  R.~F. 2006, \apj, 640, 1142

\bibitem[{{Rees} {et~al.}(2000){Rees}, {L{\'o}pez Ariste}, {Thatcher}, \&
  {Semel}}]{Rees+others2000}
{Rees}, D.~E., {L{\'o}pez Ariste}, A., {Thatcher}, J., \& {Semel}, M. 2000,
  \aap, 355, 759

\bibitem[{{Rimmele}(2000)}]{Rimmele2000}
{Rimmele}, T.~R. 2000, in Proc. SPIE Vol. 4007, p. 218-231, Adaptive Optical
  Systems Technology, Peter L. Wizinowich; Ed., ed. P.~L. {Wizinowich},
  218--231

\bibitem[{{S{\'a}nchez Almeida} \& {Lites}(2000)}]{SanchezAlmeida+Lites2000}
{S{\'a}nchez Almeida}, J. \& {Lites}, B.~W. 2000, \apj, 532, 1215

\bibitem[{{Sasso} \& {Solanki}(2006)}]{Sasso+Lagg+Solanki2006}
{Sasso}, C., Lagg, A. \& {Solanki}, S. 2006, A\&A, 456, 367

\bibitem[{{Scharmer} {et~al.}(2003){Scharmer}, {Bjelksjo}, {Korhonen},
  {Lindberg}, \& {Petterson}}]{Scharmer2003}
{Scharmer}, G.~B., {Bjelksjo}, K., {Korhonen}, T.~K., {Lindberg}, B., \&
  {Petterson}, B. 2003, in Innovative Telescopes and Instrumentation for Solar
  Astrophysics. Edited by Stephen L. Keil, Sergey V. Avakyan . Proceedings of
  the SPIE, Volume 4853, pp. 341-350 (2003)., ed. S.~L. {Keil} \& S.~V.
  {Avakyan}, 341--350

\bibitem[{{Shimizu}(2004)}]{Shimizu2004}
{Shimizu}, T. 2004, in ASP Conf. Ser. 325: The Solar-B Mission and the
  Forefront of Solar Physics, ed. T.~{Sakurai} \& T.~{Sekii}, 3--+

\bibitem[{{Socas-Navarro}(2005)}]{Socas-Navarro2005a}
{Socas-Navarro}, H. 2005, \apjl, 631, L167

\bibitem[{{Socas-Navarro} {et~al.}(2001){Socas-Navarro}, {L{\'o}pez Ariste}, \&
  {Lites}}]{Socas-Navarro+others2001}
{Socas-Navarro}, H., {L{\'o}pez Ariste}, A., \& {Lites}, B.~W. 2001, \apj, 553,
  949

\bibitem[{{Socas-Navarro} {et~al.}(2004){Socas-Navarro}, {Trujillo Bueno}, \&
  {Landi Degl'Innocenti}}]{Socas-Navarro+TrujilloBueno+LandiDeglInnocenti2004}
{Socas-Navarro}, H., {Trujillo Bueno}, J., \& {Landi Degl'Innocenti}, E. 2004,
  \apj, 612, 1175

\bibitem[{{Socas-Navarro} {et~al.}(2000{\natexlab{a}}){Socas-Navarro},
  {Trujillo Bueno}, \& {Ruiz Cobo}}]{Socas-Navarro+others2000b}
{Socas-Navarro}, H., {Trujillo Bueno}, J., \& {Ruiz Cobo}, B.
  2000{\natexlab{a}}, Science, 288, 1398

\bibitem[{{Socas-Navarro} {et~al.}(2000{\natexlab{b}}){Socas-Navarro},
  {Trujillo Bueno}, \& {Ruiz Cobo}}]{Socas-Navarro+others2000}
---. 2000{\natexlab{b}}, \apj, 530, 977

\bibitem[{{Socas-Navarro} \& {Uitenbroek}(2004)}]{Socas-Navarro+Uitenbrock2004}
{Socas-Navarro}, H. \& {Uitenbroek}, H. 2004, \apjl, 603, L129

\bibitem[{{Socas-Navarro} {et~al.}(2006)}]{Socas-Navarro+others2006a}
{Socas-Navarro}, H. {et~al.} 2006, Solar Physics, 235, 55

\bibitem[{{Stein} \& {Nordlund}(2006)}]{Stein+Nordlund2006}
{Stein}, R.~F. \& {Nordlund}, {\AA}. 2006, \apj, 642, 1246

\bibitem[{{Uitenbroek} {et~al.}(2006){Uitenbroek}, {Balasubramaniam}, \&
  {Tritschler}}]{Uitenbroek+others2006}
{Uitenbroek}, H., {Balasubramaniam}, K.~S., \& {Tritschler}, A. 2006, \apj,
  645, 776

\bibitem[{{Vecchio} {et~al.}(2006)}]{Vecchio+others2006} {Vecchio}, A., {Cauzzi}, G., {Reardon}, K.~P., {Janssen}, K. \& {Rimmele}, T. 2006, astro-ph/0611206

\bibitem[{{Wikst{\o}l} {et~al.}(2000){Wikst{\o}l}, Hansteen, Carlsson, \&
  Judge}]{Wikstol+others2000}
{Wikst{\o}l}, {\O}., Hansteen, V., Carlsson, M., \& Judge, P.~G. 2000,
  Astrophys.\ J., 531, 1150

\bibitem[{{Zhang} \& {Zhang}(2000)}]{Zhang+Zhang2000}
{Zhang}, H. \& {Zhang}, M. 2000, \solphys, 196, 269


\end{thebibliography}

\end{document}